\documentclass[12pt, a4paper]{article}
\usepackage{amsmath,amssymb,amsfonts}
\usepackage{graphicx}
\usepackage[colorlinks]{hyperref}

\renewcommand{\theequation}{\arabic{section}.\arabic{equation}}

\begin{document}

\title{Induced vacuum magnetic field in \\
the cosmic string background}

\author{Yurii A. Sitenko}

\date{}

\maketitle
\begin{center}
Bogolyubov Institute for Theoretical Physics,\\
National Academy of Sciences of Ukraine,\\
14-b Metrologichna Street, 03143 Kyiv, Ukraine

\end{center}

\begin{abstract}
The relativistic charged spinor matter field is quantized in the background of a straight cosmic string with nonvanishing transverse size. The most general boundary conditions ensuring the impossibility for matter to penetrate through the edge of the string core are considered. The role of discrete symmetries is elucidated, and analytic expressions for the temporal and spatial components of the induced vacuum current are derived in the case of either $P$ or $CT$ invariant boundary condition with two parameters varying arbitrarily from point to point of the edge. The requirement of physical plausibility for the global induced vacuum characteristics is shown to remove completely an arbitrariness in boundary conditions. We find out that a magnetic field is induced in the vacuum and that a sheath in the form of a tube of the magnetic flux lines encloses a cosmic string. The dependence of the induced vacuum magnetic field strength on the string flux and tension, as well as on the transverse size of the string and on the distance from the string, is unambiguously determined. 


\end{abstract}

\section{Introduction}
\setcounter{equation}{0}

According to the standard cosmological model, the early universe, in the course of its expansion and cooling, undergoes phase transitions with spontaneous symmetry breaking. Topological defects of various kinds (monopoles, strings, domain walls) are formed in the aftermath of such phase transitions \cite{Ki1,Ki2,Zel,Vil1,Vil2}. Most of these defects are unstable and decay as the universe expands, but some can survive. Namely, linear defects (cosmic strings), starting from a random tangle, evolve into two different sets: the unstable one which consists of a variety of string loops decaying by gravitational radiation and the stable one which consists of several long, approximately straight strings spanning the horizon, see, e.g., reviews in \cite{Vil3,Ki3}. Although a measurement of the cosmic microwave background radiation testifies that the cosmic strings are not abundant, their evolution brings distinct astrophysical effects, in particular, producing detectable gravitational waves \cite{Dam}, gamma-ray bursts \cite{Ber}, and high-energy cosmic rays \cite{Bhat}. The interest in cosmic strings also is amplified by theoretical findings that they almost inevitably emerge in the framework of superstring and supergravity models aiming at the explanation of the inflation (brane inflation)\cite{Sar,Jean,Dva,Pol,Sak,Cop}. 

A straight infinitely long cosmic string in its rest frame is characterized by tension (or linear density of energy)
\begin{equation}\label{1.1}
\mu=\frac{1}{16 \pi \c{G}} \int\limits_{\rm core}{\rm d}\sigma\,R,
\end{equation}
where $\c{G}$ is the gravitational constant, $R$ is the curvature scalar, and the integration is over the transverse section of the string core 
(natural units $\hbar=c=1$ are used). Space outside the string core is locally flat ($R=0$) but
non-Euclidean, with squared length element
\begin{equation}\label{1.2}
ds^2= dr^2+(1-4\c{G} \mu)^2 r^2 d\varphi^2+dz^2,
\end{equation}
where cylindrical coordinates with the symmetry axis coinciding with the axis of the string are chosen. Such a space can be denoted as a conical one: a surface which is transverse to the string is isometric to the surface of a cone with the deficit angle equal to $8\pi \c{G}\mu$. The observation of discontinuities in the cosmic microwave background radiation imposes an upper bound on the value of the deficit angle to be $2.5 \times 10^{-6}$ rad, see \cite{Bat}; however, it makes sense in a more general context in theory to consider larger values of the deficit angle up to $2\pi$. Moreover, of some interest may be saddle-like conical spaces with the deficit angle taking negative values unbounded from below (unbounded surplus angle), although tension \eqref{1.1} is negative in this case; for instance, such a space effectively emerges in the A phase of superconductive He3 (see \cite{Vol}).
Various nanoconical structures arise in a diverse set of condensed matter systems known as the Dirac materials, ranging from honeycomb crystalline allotropes (graphene \cite{Ge}, silicene and germanene \cite{Cah}, phosphorene \cite{Liu}) to high-temperature cuprate superconductors \cite{Tsu} and topological
insulators \cite{Qi}.

Returning to the concept of a string as of a topological defect, we note that, in the case of spontaneous breakdown of a continuous symmetry, the string  acquires an additional global characteristic  -- flux
\begin{equation}\label{1.3}
\Phi=\int\limits_{\rm core} d\sigma\, \mbox{\boldmath $\partial$}\times \textbf{V}=\oint d \textbf{x} \cdot \textbf{V},
\end{equation}
where $\textbf{V}$ is the vector potential of the gauge field corresponding to the spontaneously broken gauge symmetry, and the line integral is over a closed contour encircling the string core once. Note also that the gauge field potential is nonvanishing outside the string core, although the gauge field strength vanishes there (this is a general ground for the renowned Aharonov-Bohm effect \cite{Ehre,Aha}).

Matter emerging in the universe in an epoch after the birth of a cosmic string is assumed to interact with its gauge field in the minimal way, i.e., via substitution 
\begin{equation}\label{1.33}
\mbox{\boldmath
$\partial$} \rightarrow \mbox{\boldmath
$\partial$} - {\rm i}\tilde e\, \textbf{V}\end{equation} 
for the gradient of the matter field ($\tilde e$ is the appropriate coupling constant). A plausible hierarchy of energy scales is as follows 
\begin{equation}\label{1.4}
 m_{{\rm Planck}} \gg m_{{\rm H}} \gg m,    
\end{equation}
where $m_{{\rm Planck}} = \c{G}^{-1/2}$ is the Planck energy scale, $m_{{\rm H}}$ is the energy scale of the spontaneous symmetry breakdown, i.e., the mass of the appropriate Higgs boson, and $m$ is the mass of the matter field. In view of the right inequality in \eqref{1.4}, the direct interaction of the matter field with the Higgs field is negligible. Assuming without a loss of generality that the string has the form of a tube of radius $r_0$, we note that the transverse size of the string core is of the order of the correlation length,
\begin{equation}\label{1.5}
r_0 \sim m_{{\rm H}}^{-1}.
\end{equation}
The string tension is of the order of the inverse correlation length squared,
\begin{equation}\label{1.6}
\mu \sim m_{{\rm H}}^2.
\end{equation}

A study of various effects of vacuum polarization of the quantum relativistic charged spinor matter field in the cosmic string background has a long history of more than three decades (see \cite{Gor,Si0,Flek,Par,Sit6,Sit9,Bez1,Bel1,Bez2,Bel2,Bez4,Bez5,SiG} and references therein). However, this study in no way is exhaustive, and several crucial points should be stressed. A simplifying asssumption in \cite{Gor,Si0,Flek,Par,Sit6,Sit9,Bel2} is to neglect the transverse size of the string. In view of \eqref{1.5}, this corresponds to a requirement of the vanishing correlation length, i.e. infinite $m_{{\rm H}}$ and, surely, infinite $m_{{\rm Planck}}$, keeping ratio $m_{{\rm H}}/m_{{\rm Planck}}$ to be small enough in order to satisfy the left inequality in \eqref{1.4}. Moreover, the matter field has to be subject to a certain (disputable) condition at the location of an infinitely thin string. If the transverse size of the string is taken into account, then an unavoidable obstruction arises so far as the interior of the string core is a ``black box'': the exact forms of the local characteristics (curvature scalar and gauge field strength) inside the core are unknown, and the results are obtained for certain made-up configurations only, see  \cite{Bez4,Bez5}. On the other hand, it should be recalled that a phase with the spontaneously broken symmetry exists outside the string core and the vacuum has to be defined there. Hence the matter field is not allowed to penetrate inside the string core, obeying a boundary condition at its edge. What in general can be said about this boundary condition? First, there is a  requirement of mathematical consistency: as long as the case of stable matter is considered, the Hamiltonian operator for the matter field has to be self-adjoint. Secondly, a physical requirement, as has been just noted, is in the absence of the penetration of the matter field through the boundary. Actually, for a spatial region with a one-component boundary (and this is space out of the string core), the requirements are equivalent, and they yield, as an outcome, a four-parameter family of boundary conditions, see \cite{Bee,Wie,Si1}; the parameter values may in general differ for different points of the boundary \cite{SiY}. One can comprehend that an arbitrariness in the configuration of the curvature scalar and the gauge field strength inside the string core is translated into an arbitrariness in the choice of boundary conditions at the edge of the string core. Thus, the quantum effects of matter in the cosmic string background depend on the boundary parameters, as well as on the global characteristics of the string -- tension \eqref{1.1} and flux \eqref{1.3}; moreover, as a manifestation of the Aharonov-Bohm effect, the dependence on the flux is periodic with a period equal to the appropriate London flux quantum, $2\pi /\tilde e$.

However, it is needless to say that the above generality is excessive, and some additional physical arguments can restrict the arbitrariness of boundary conditions. For instance, a requirement of various discrete symmetries reduces the number of boundary parameters. As will be shown in the present study, there are further physical requirements that completely remove the arbitrariness of boundary conditions.        

Observation of magnetic fields of order  $10^{-18} - 10^{-10}$ Gauss in intergalactic voids \cite{Nero} indicates that a magnetic field has emerged in the early universe. The generation of primordial magnetic fields is a hot topic, and different scenarios for this phenomenon are proposed, see \cite{Subr}. Therefore, among a whole variety of the vacuum polarization effects  in the cosmic string background, we focus on the induced vacuum magnetic field. As is proven in the present paper, the cosmic string background indeed produces a magnetic field in the vacuum of the quantum relativistic charged spinor matter field, and the dependence of this effect on the global characteristics of the string is unambiguously determined.

In the next section we define the physical characteristics of the vacuum of the quantum relativistic charged spinor matter field in the cosmic string background . In Section 3 we choose boundary conditions ensuring the absence of the matter flux through the edge of the string core and display the role of discrete symmetries. The induced vacuum current and magnetic field strength in the case of the two-parameter position-dependent boundary condition are obtained in Section 4. In Section 5 we consider the total induced vacuum magnetic flux and find out that an arbitrariness in the boundary parameter values is removed by requiring the physically plausible behavior for the flux. Finally, the results are summarized in Section 6. In Appendix A we present the complete set of solutions to the Dirac equation that is relevant to the problem considered. An alternative representation of the results, allowing for the explicit extraction of the dependence on the transverse size of the string, is given in Appendix B. The results for the formal case of the infinitely thin string are given in Appendix C.

\section{Vacuum of quantum spinor matter in \\
the cosmic string background}
\setcounter{equation}{0}

The current that is induced in the vacuum by static background fields is defined by relations
\begin{multline}\label{2.1}
j^0(\textbf{x})= \frac12 \langle {\rm vac}| \left[{\hat\Psi}^\dag(\textbf{x},x^0) \, {\hat\Psi}(\textbf{x},x^0) - {\hat\Psi}^T(\textbf{x},x^0) \, {\hat\Psi}^{\dag T}(\textbf{x},x^0)\right] |{\rm vac}
\rangle \\
=-\frac12\sum\hspace{-1.4em}\int \rm{sgn}(E) \, 
\psi^\dag_E(\textbf{x}) \, \psi_E(\textbf{x})
\end{multline}
and
\begin{multline}\label{2.2}
\textbf{j}(\textbf{x})= \frac12 \langle {\rm vac}| \left[{\hat\Psi}^\dag(\textbf{x},x^0) \, \gamma^0
\mbox{\boldmath $\gamma$} \, {\hat\Psi}(\textbf{x},x^0) - {\hat\Psi}^T(\textbf{x},x^0)\left(\gamma^0
\mbox{\boldmath $\gamma$}\right)^T {\hat\Psi}^{\dag T}(\textbf{x},x^0)\right] |{\rm vac}
\rangle \\
=-\frac12\sum\hspace{-1.4em}\int \rm{sgn}(E) \, 
\psi^\dag_E(\textbf{x}) \, \gamma^0
\mbox{\boldmath $\gamma$} \, \psi_E(\textbf{x})
\end{multline}
[$\rm{sgn}(t)$ is the sign function, $\rm{sgn}(t)=\pm 1$ at $t \gtrless 0$]; here 
\begin{equation}\label{2.3}
 {\hat \Psi}({\textbf{x}},x^0)=\sum\hspace{-1.6em}\int\limits_{E>0} {\rm e}^{-{\rm
i}E x^0}\psi_{E}({\bf x})\,{\hat a}_{E}+
 \sum\hspace{-1.6em}\int\limits_{E<0}
  {\rm e}^{-{\rm i}E x^0}\psi_{E}({\bf
  x})\,{\hat b}^\dag_{E}
\end{equation}
is the operator of the second-quantized spinor field, superscripts $T$ and $\dag$ denote a transposition and a  Hermitian conjugation, ${\hat a}^\dag_E $ and ${\hat a}_E$ (${\hat b}^\dag_E $ and ${\hat b}_E$) are the spinor particle (antiparticle) creation and destruction operators obeying the anticommutation relations, ground state $|{\rm vac} \rangle$ is defined by relation ${\hat a}_E|{\rm vac}\rangle = {\hat b}_E|{\rm vac}\rangle = 0$, symbol \mbox{$\displaystyle \sum\hspace{-1.4em}\int $} denotes summation over the discrete part and integration (with a certain measure) over the continuous part of the energy spectrum, and ${\rm e}^{-{\rm i}E x^0}\psi_{E}({\bf x})$ is the solution to the stationary equation with the appropriate Dirac Hamiltonian denoted by $H$,
\begin{equation}\label{2.4}
 \left({\rm i}\partial_0 - H\right) {\rm e}^{-{\rm i}E x^0}\psi_{E}({\bf x}) = 0.
\end{equation}
If the current given by \eqref{2.1} and \eqref{2.2} is nonvanishing, then electric [$\textbf{E}_I(\textbf{x})$] and magnetic [$\textbf{B}_I(\textbf{x})$] field strengths are also induced in the vacuum, as a consequence of the Maxwell equations,
\begin{equation}\label{2.5}
\mbox{\boldmath $\partial$} \cdot \textbf{E}_I(\textbf{x}) =
e\, j^0(\textbf{x})
\end{equation}
and
\begin{equation}\label{2.6}
\mbox{\boldmath $\partial$}\times \textbf{B}_I(\textbf{x}) =
e\, \textbf{j}(\textbf{x}),
\end{equation}
where the electromagnetic coupling constant, $e$, differs in general from $\tilde e$. The total flux of the induced vacuum magnetic field is
\begin{equation}\label{2.7}
\Phi_I=\int d \, \textbf{S} \cdot \textbf{B}_I(\textbf{x}),
\end{equation}
while the global characteristics that is related to 
$\textbf{E}_I(\textbf{x})$ is the total induced vacuum charge,
\begin{equation}\label{2.8}
Q_{\rm I} = \int d \, V \mbox{\boldmath $\partial$} \cdot \textbf{E}_I(\textbf{x});
\end{equation}
here $d \, V$ is the infinitesimal element of the volume of space, and $d \, \textbf{S}$ is the infinitesimal element of the surface which is orthogonal to $\textbf{B}_I(\textbf{x})$.

In the case of the cosmic string background, the Dirac Hamiltonian takes form
\begin{equation}\label{2.9}
H=-{\rm i} \gamma^0
\mbox{\boldmath $\gamma$}\cdot \left[\mbox{\boldmath
$\partial$} - {\rm i}\tilde e\, \textbf{V}(\textbf{x}) + \frac{{\rm i}}2
\mbox{\boldmath $\omega$}(\textbf{x})\right]+\gamma^0 m,
\end{equation}
where $\textbf{V}(\textbf{x})$ is the bundle connection and $\mbox{\boldmath $\omega$}(\textbf{x})$ is the spin connection, corresponding to the cosmic string background. In cylindrical coordinates $r, \varphi, z$ with the symmetry axis coinciding with the axis of a straight cosmic string, only angular components of the connections are nonvanishing,
\begin{equation}\label{2.10}
    V_\varphi=\frac{\Phi}{2\pi}, \quad w_\varphi={\rm
    i}\frac{1 -\nu}r\, \gamma_\varphi \gamma^r,
\end{equation}
and Hamiltonian \eqref{2.9} takes form
\begin{equation}\label{2.11}
H=-{\rm i}\gamma^0
\left[\gamma^r\left(\partial_r+\frac{1-\nu}{2r}\right)+\gamma^\varphi\left(\partial_\varphi-{\rm
i}\frac{\tilde e \Phi}{2\pi}\right)+\gamma^3 \, \partial_z\right]+\gamma^0 m,
\end{equation}
where $\Phi$ is given by \eqref{1.3}, notation
\begin{equation}\label{2.111}
\nu=(1-4\c{G} \mu)^{-1}
\end{equation}
is introduced, and $\mu$ is given by \eqref{1.1}. Using the following representation for the Dirac matrices, 
\begin{equation}\label{2.12}
\gamma^0=\left(\begin{array}{cc}
                    \sigma^3 & 0\\
                   0 & \sigma^3
                     \end{array}\right),\quad
\gamma^1={\rm i}\left(\begin{array}{cc}
                    \sigma^1 & 0\\
                    0 & \sigma^1
                     \end{array}\right),\quad
\gamma^2={\rm i}\left(\begin{array}{cc}
                    \sigma^2 & 0\\
                    0 & - \sigma^2
                     \end{array}\right),\quad
\gamma^3={\rm i}\left(\begin{array}{cc}
                    0 & \sigma^2\\
                    \sigma^2 & 0
                     \end{array}\right)
\end{equation}
($\sigma^1$, $\sigma^2$, and $\sigma^3$ are the Pauli matrices), we obtain the block-diagonal form for matrices 
$\gamma^r$ and $\gamma^\varphi$,
\begin{equation}\label{2.13}
\gamma^r=\gamma_r={\rm i}
\left(\begin{array}{cc}
                    \sigma^1\cos\varphi + \sigma^2\sin\varphi& 0 \\
                    0 & \sigma^1\cos\varphi - \sigma^2\sin\varphi
                     \end{array}\right)
\end{equation}
and 
\begin{equation}\label{2.14}
\gamma^\varphi=\frac{\nu}{r}\,{\rm i}
\left(\begin{array}{cc}
                    \sigma^2\cos\varphi - \sigma^1\sin\varphi& 0 \\
                    0 & - \sigma^2\cos\varphi - \sigma^1\sin\varphi
                     \end{array}\right),
\quad \gamma_\varphi = \frac{r^2}{\nu^2} \, \gamma^\varphi,
\end{equation}
and present Hamiltonian \eqref{2.11} as
\begin{equation}\label{2.15}
H=\left(\begin{array}{cc}
                    H_1 & - {\rm i}\sigma^1 \partial_z\\
                    - {\rm i}\sigma^1 \partial_z & H_{-1}
                     \end{array}\right),
\end{equation}
where
\begin{multline}\label{2.16}
H_s=-{\rm i}\left[\left(s \sigma^1\sin\varphi - \sigma^2\cos\varphi\right)\left(\partial_r+\frac{1-\nu}{2r}\right) \right.\\
\left. + \frac{\nu}{r}\left(s \sigma^1\cos\varphi  + \sigma^2\sin\varphi\right)\left(\partial_\varphi-{\rm
i}\frac{\tilde e \Phi}{2\pi}\right)\right]+\sigma^3 m, \quad s= \pm 1.
\end{multline}
Decomposing the four-component function, $\psi_E(r,\varphi, z)$, into the two-component ones, $\psi_E^{(1)}(r,\varphi)$  and $\psi_E^{(-1)}(r,\varphi)$,
\begin{equation}\label{2.17}
\psi_E(r,\varphi, z) =\frac{e^{{\rm i}k_3 z}}{\sqrt{2\pi}}                    \left(\begin{array}{c}
                   \psi_E^{(1)}(r,\varphi) \\
                   {\rm i} \psi_E^{(-1)}(r,\varphi)
                    \end{array}\right),
\end{equation}
we note that the stationary Dirac equation, see \eqref{2.4}, 
\begin{equation}\label{2.18}
\left(E - H\right)\psi_E(r,\varphi, z) = 0,
\end{equation}
is equivalent to system of equations
\begin{equation}\label{2.19}
 \left\{
 \begin{array}{c}
\left(E - H_{1}\right) \psi_E^{(1)}(r,\varphi) = {\rm i} k_3 \sigma^1 \psi_E^{(-1)}(r,\varphi) \\
\left(E - H_{-1}\right) \psi_E^{(-1)}(r,\varphi) = - {\rm i} k_3 \sigma^1 \psi_E^{(1)}(r,\varphi)
\end{array}
\right\}.
\end{equation}
In view of equalities
\begin{equation}\label{2.20}
H_{1} \sigma^1 + \sigma^1 H_{-1} = \sigma^1 H_{1} + H_{-1} \sigma^1 = 0,
\end{equation}
relation
\begin{equation}\label{2.21}
\left(E^2 - H^2\right)\psi_E(r,\varphi, z) = 0
\end{equation}
which is just a direct consequence of \eqref{2.18} is equivalent to relation
\begin{equation}\label{2.22}
\left(E^2 - k^2_3 - H^2_s\right)\psi_E^{(s)}(r,\varphi) = 0.
\end{equation}
Thus, function $\psi_E^{(s)}(r,\varphi)$ can be regarded as a solution to equation
\begin{equation}\label{2.23}
\left(E - \left.H_s\right|_{m \rightarrow m_3}\right)\psi_E^{(s)}(r,\varphi) = 0, \quad m_3 = \sqrt{m^2 + k^2_3}.
\end{equation}
Decomposing function $\psi_E^{(s)}(r,\varphi)$ as
\begin{equation}\label{2.24}
\psi_E^{(s)}(r,\varphi) = \sum_{n \in \mathbb{Z}}
                   \left(\begin{array}{c}
                   f_n^{(s)}(r,E ) \, \exp\left[{\rm i}\left(n + \frac12 - \frac12 \, s\right)\varphi\right] \\
                   g_n^{(s)}(r,E ) \, \exp\left[{\rm i}\left(n + \frac12 + \frac12 \, s\right)\varphi\right]
                    \end{array}\right)
\end{equation}
($\mathbb{Z}$ is the set of integer numbers), we present the latter equation as a system of two first-order differential equations for radial functions:
\begin{equation}\label{2.25}
 \left\{
 \begin{array}{c}
\left\{-\partial_r + r^{-1} \left[s \nu \left(n-n_{\rm c}\right)-G_s\right]\right\} f_n^{(s)}(r,E) =\left(E+m_3\right) g_n^{(s)}(r,E) \\
\left\{\partial_r + r^{-1} \left[s \nu \left(n-n_{\rm c}\right)+1-G_s\right]\right\} g_n^{(s)}(r,E)
=\left(E-m_3\right) f_n^{(s)}(r,E)
\end{array}
\right\},
\end{equation}
where
\begin{equation}\label{2.26}
n_{\rm c}=\left[\!\left| \frac{\tilde e \Phi}{2\pi}
\right|\!\right], \quad F = \left\{\!\!\left| \frac{\tilde
e \Phi}{2\pi}  \right|\!\!\right\},\quad G_s =s \nu \left(F  - \frac12 \right)+ \frac12,
\end{equation}
$\left[\!\left| u \right|\!\right]$ is the integer part of quantity $u$ (i.e., the integer that
is less than or equal to $u$), and $ \left\{\!\!\left|  u \right|\!\!\right\} = u - \left[\!\left| u \right|\!\right]$ is the fractional part of quantity $u$, $ 0\leq  \left\{\!\!\left| u \right|\!\!\right\}<1 $.

Using \eqref{2.12} and \eqref{2.24}, one immediately gets $j_r =0$ and $j_z =0$, while the remaining components of the induced vacuum current are $r$-dependent:
\begin{equation}\label{2.27}
j^0(r) = - \frac 12 \sum_{s = \pm 1}\sum\hspace{-1.4em}\int \sum_{n \in
\mathbb{Z}} {\rm sgn} (E) \left\{\left[f_n^{(s)}(r,E)\right]^2 + \left[g_n^{(s)}(r,E)\right]^2\right\}
\end{equation}
and
\begin{equation}\label{2.28}
j_\varphi(r) = - \frac r\nu \sum_{s = \pm 1}\sum\hspace{-1.4em}\int \sum_{n \in
\mathbb{Z}} {\rm sgn} (E) \, s \, f_n^{(s)}(r,E) \, g_n^{(s)}(r,E).
\end{equation}
As is noted in Introduction, parameters of the boundary condition in general case vary from point to point of the boundary. Consequently, $j^0(r)$ \eqref{2.27} and $j_\varphi(r)$ \eqref{2.28} in this case contain an additional dependence on 
$\varphi$ and $z$ owing to the boundary condition. A variation of the boundary parameters with $z$ can be moderate enough, and a violation of the translational invariance along the string axis can be regarded as negligible. Therefore, the induced vacuum magnetic field strength is directed along the string axis,
\begin{equation}\label{2.30}
B_{\rm I}(r) = e \nu \int\limits_r^\infty \frac{dr'}{r'} \,
j_\varphi(r'),
\end{equation}
with total flux
\begin{equation}\label{2.31}
\Phi_{\rm I} = \frac{1}{\nu} \int\limits_{0}^{2\pi} d \varphi \int\limits_{r_0}^\infty dr\, r B_{\rm
I}(r)
\end{equation}
($r_0$ is the radius of the string). The total induced vacuum charge is given by expression
\begin{equation}\label{2.29}
Q_{\rm I} = \frac{e}{\nu} \int\limits_{0}^{2\pi} d \varphi \int\limits_{ - \infty}^\infty dz \int\limits_{r_0}^\infty dr\, r j^0(r).
\end{equation} 
In the following we shall find quantities \eqref{2.27}-\eqref{2.29} in the case of $1 \leq \nu < \infty$, which corresponds to the deficit angle ranging from $0$ to $2\pi$,   and in the case of $\frac12 \leq \nu < 1$, which corresponds to the deficit angle ranging from $-2\pi$ to $0$, for $\frac12 \left(\frac1\nu - 1\right) < F < \frac12 \left(3 - \frac1\nu\right)$.

\section{Boundary condition: self-adjointness and discrete symmetries}

\setcounter{equation}{0}

The most general boundary condition that is compatible with the self-adjointness of the Dirac Hamiltonian in the case of three-dimensional spatial region $X$ with one-component boundary $\partial{X}$ depends on four parameters $u,v,t^1,t^2$, see \cite{Si1}:
\begin{equation}\label{3.1}
(I - K)\psi|_{\mathbf{x}\in
\partial{X}} = 0
\end{equation}
with
\begin{equation}\label{3.2}
K=\frac{(1+u^2-v^2-{\boldsymbol{t}}^2)I+(1-u^2+v^2+{\boldsymbol{t}}^2)\gamma^0}{2{\rm
i}(u^2-v^2-{\boldsymbol{t}}^2)}(u\boldsymbol{n}\cdot\boldsymbol{\gamma}+v\gamma^{5}-{\rm
i}\boldsymbol{t}\cdot\boldsymbol{\gamma});
\end{equation}
here $\gamma^{5}=-{\rm i}\gamma^0\gamma^1\gamma^2\gamma^3$, 
$\boldsymbol{n}$ is the outward normal to $\partial{X}$, 
${\boldsymbol{n}}^2=1$, and
$\boldsymbol{t}=(t^1,t^2)$ is tangential to $\partial{X}$, $\boldsymbol{t}\cdot\boldsymbol{n}=0$. It should be recalled that  
\begin{equation}\label{3.3}
K = - {\rm i} \boldsymbol{n}\cdot\boldsymbol{\gamma}
\end{equation}
corresponds to the so-called quark bag boundary condition that was postulated a long time ago as the condition ensuring the confinement of the matter field, see \cite{Bog,Cho1,Joh}. However, the confinement is ensured equally as well by the four-parameter boundary condition with matrix $K$ given by \eqref{3.2}, and this is the most general confining boundary condition. 
 
Parameters of the boundary condition, $u,v,t^1,t^2$,
can be interpreted as the self-adjoint extension parameters. It should be emphasized that the values 
of these parameters may in general vary from point to point of the
boundary. In this respect the ``number'' of self-adjoint extension
parameters is, in fact, infinite, moreover, it is not countable but
is of power of a continuum. This distinguishes the case of an
extended boundary from the case of an excluded point (contact
interaction), when the number of self-adjoint extension parameters
is finite, being equal to $n^2$ for the deficiency index equal to
($n,n$) (see, e.g., \cite{Ree,Alb}). 

In the case of spatial region out of the straight cosmic string of radius $r_0$, the boundary condition is 
\begin{equation}\label{3.4}
(I-K)\left.\psi \right|_{r = r_0} = 0 
\end{equation}
with
\begin{multline}\label{3.5}
K= \frac{\left[1+u^2-v^2-(\nu/r)^2(t_\varphi)^2-(t^z)^2\right]I+\left[u^2+v^2+(\nu/r)^2(t_\varphi)^2+(t^z)^2\right]\gamma^0}{2\left[u^2-v^2-(\nu/r)^2(t_\varphi)^2-(t^z)^2\right]}  \\
\times\left({\rm i}u\gamma^r-{\rm i}v\gamma^5-t_\varphi\gamma^\varphi-t^z\gamma^z\right).
\end{multline}
Invariance under spatial reflection, 
\begin{equation}\label{3.6}
P:\ \ \ \varphi\rightarrow\varphi+\pi\ \ \ z\rightarrow -z\ \ \ \psi\rightarrow {\rm i}\gamma^0\psi, 
\end{equation}
reduces the number of boundary parameters by half:
\begin{equation}\label{3.7}
v=0,\ \ t^z=0,
\end{equation}
and matrix $K$ takes form
\begin{equation}\label{3.8}
K=\frac{\left[1+u^2-(\nu/r)^2(t_\varphi)^2\right]I+\left[1-u^2+(\nu/r)^2(t_\varphi)^2\right]\gamma^0}{2\left[u^2-(\nu/r)^2(t_\varphi)^2\right]}({\rm i}u\gamma^r-t_\varphi\gamma^\varphi).
\end{equation}
Using parametrization
\begin{equation}\label{3.9}
u=\frac{\cos \eta}{\cos \theta+\sin \eta},\ \ t_\varphi=\frac{r}{\nu}\,\frac{\sin \theta}{\cos \theta+\sin \eta},
\end{equation}
we get
\begin{equation}\label{3.10}
K=\frac{I\cos \theta+\gamma^0\sin\eta}{\cos^2\theta-\sin^2\eta}\left({\rm i}\gamma^r\cos\eta-\frac{r}{\nu}\gamma^\varphi\sin\theta\right).
\end{equation}
In view of invariance under the simultaneous shift by $\pi$,
\begin{equation}\label{3.11}
\left.K\right|^{\theta\rightarrow\theta+\pi}_{\eta\rightarrow\eta+\pi} = K,
\end{equation}
it suffices to restrict the range of boundary angular parameters to
\begin{equation}\label{3.12}
0\leq \theta<2\pi, \,\,\,\, 0\leq\eta<\pi.
\end{equation}
A dependence of parameters $\theta$ and 
$\eta$ on $\varphi$ and $z$ is admissible, if condition 
\begin{equation}\label{3.122}
 \theta(\varphi+\pi, -z) = \theta(\varphi, z), \,\,\,\, \eta(\varphi+\pi, -z) = \eta(\varphi, z)
\end{equation}
is satisfied.

It should be noted (see \cite{Si3} for details) that boundary condition \eqref{3.1} can be rewritten as 
\begin{equation}\label{3.13}
(I-\tilde{K})\left.\psi\right|_{\mathbf{x}\in
\partial{X}}=0,
\end{equation}
where
\begin{equation}\label{3.14}
\tilde{K}=(1-N)K+N.
\end{equation}
If $N$ is a Hermitian matrix, $N^{\dagger}=N$,  which obeys condition 
\begin{equation}\label{3.15}
N K - K^{\dagger} N = K - K^\dagger ,
\end{equation}
then $\tilde{K}$ is Hermitian also, $\tilde{K}^{\dagger}=\tilde{K}$. In the case of $K$ given by \eqref{3.10}, we have 
\begin{equation}\label{3.16}
N=\gamma^0\cos\theta\sin \eta+{\rm i}\frac{r}{\nu}\gamma^r\gamma^\varphi\sin\theta\cos\eta
\end{equation}
and
\begin{equation}\label{3.17}
\tilde{K}=\left(I\cos\theta-\frac{r}{\nu}\gamma^\varphi\sin\theta\right)
\left({\rm i}\gamma^r\cos\eta+\gamma^0\sin\eta\right).
\end{equation}
Note the following relation under separate shifts by $\pi$:
\begin{equation}\label{3.18}
\left.\tilde{K}\right|_{\theta\rightarrow\theta+\pi}=\left.\tilde{K}\right|_{\eta\rightarrow\eta+\pi}=-\tilde{K}.
\end{equation}

Imposing the boundary condition with either $K$ \eqref{3.10} or $\tilde{K}$ \eqref{3.17} on the decomposition of the wave function into modes, see \eqref{2.17} and \eqref{2.24}, we obtain  condition for the modes:
\begin{equation}\label{3.19}
\left.f_n^{(s)}\right|_{r=r_0}+\tau_s(\theta,\eta)\left.g_n^{(s)}\right|_{r=r_0}=0,
\end{equation}
where
\begin{equation}\label{3.20}
\tau_s(\theta,\eta)=\tan\left(s\frac{\theta}{2}+\frac{\eta}{2}+\frac{\pi}{4}\right).
\end{equation}
To end the discussion of the $P$ invariant boundary condition, we list some properties of $\tau_s(\theta,\eta)$:
\begin{multline}\label{3.21}
\tau_s(-\theta,-\eta)=\tau_s^{-1}(\theta,\eta), \, \tau_{-s}(\theta,\eta)=\tau_s^{-1}(\theta,-\eta), \, 
\tau_{s}(\theta+\pi,\eta)=\tau_{s}(\theta,\eta+\pi)=-\tau_s^{-1}(\theta,\eta).
\end{multline}

The number of boundary parameters is reduced by half as well, if, instead of $P$ invariance, one requires invariance under the combination of charge conjugation and time reversal,
\begin{equation}\label{3.22}
CT:\,\,\,\,\,\,\,x^0\rightarrow-x^0\,\,\,\,\,\,\,\,\psi\rightarrow-{\rm i}\gamma^0\gamma^5\psi.
\end{equation}
Then
\begin{equation}\label{3.23}
v=0,\,\,\,\,u^2-(\nu/r)^2(t_\varphi)^2-(t^z)^2=1,
\end{equation}
and matrix $K$ takes form
\begin{equation}\label{3.24}
K={\rm i}\gamma^r\sec\theta-\left(\frac{r}{\nu}\gamma^\varphi\cos\zeta+\gamma^3\sin\zeta\right)\tan\theta,
\end{equation}
where the use is made of parametrization
\begin{equation}\label{3.25}
u=\sec\theta,\,\,\,\,t_\varphi=\frac r\nu\tan\theta\cos\zeta,\,\,\,\,t^z=\tan\theta\sin\zeta;
\end{equation}
it suffices to restrict the range to 
\begin{equation}\label{3.26}
0\leq\theta<2\pi,\,\,\,\,0\leq\zeta<\pi,
\end{equation}
and a dependence of $\theta$ and $\zeta$ on $\varphi$ and $z$ is admissible.
Determining Hermitian matrix $N$ obeying condition \eqref{3.15}, and defining Hermitian matrix $\tilde{K}$ by \eqref{3.14}, we get 
\begin{equation}\label{3.27}
N={\rm i}\gamma^r\left(\frac{r}{\nu}\gamma^{\varphi}\cos\zeta+\gamma^3\sin\zeta\right)\sin\theta
\end{equation}
and
\begin{equation}\label{3.28}
\tilde{K}={\rm i}\gamma^r\left[I\cos\theta+\left(\frac{r}{\nu}\gamma^\varphi\cos\zeta+\gamma^3\sin\zeta\right)\sin\theta\right].
\end{equation}
Because of the appearance of $\gamma^3$ in \eqref{3.24} and \eqref{3.28}, the boundary condition mixes the components of the wave function with $s=1$ and $s=-1$. Assuming nevertheless the condition for the modes in the form, cf. \eqref{3.19},
\begin{equation}\label{3.29}
\left.f_n^{(s)}\right|_{r=r_0}+\tau_s(\theta,\zeta)\left.g_n^{(s)}\right|_{r=r_0}=0,
\end{equation}
one gets relation
\begin{equation}\label{3.30}
\tau_s(\theta,\zeta)\left(1-s\sin\theta\cos\zeta\right)+\tau_{-s}(\theta,\zeta)(1+s\sin\theta\cos\zeta)=\cos\theta\left[1+\tau_s(\theta,\zeta)\tau_{-s}(\theta,\zeta)\right].
\end{equation}

If one requires $P$ invariance \eqref{3.6} in addition to $CT$ invariance \eqref{3.22}, then
\begin{equation}\label{3.31}
t^z=0,
\end{equation}
and one obtains
\begin{equation}\label{3.32}
K={\rm i}\gamma^r\sec\theta-\frac{r}{\nu}\gamma^\varphi\tan\theta
\end{equation}
and
\begin{equation}\label{3.33}
\tilde{K}={\rm i}\gamma^r\left(I\cos\theta+\frac{r}{\nu}\gamma^\varphi\sin\theta\right);
\end{equation}
the same is certainly obtained if one reverses the order of requirements. Thus, in the case of invariance under the $CPT$ transformation, 
\begin{equation}\label{3.34}
CPT:\,\,\,\, \varphi\rightarrow\varphi+\pi\,\,\,\,z\rightarrow -z\,\,\,\,x^0\rightarrow -x^0\,\,\,\,\psi\rightarrow \gamma^5\psi,
\end{equation}
the boundary condition depends on one parameter, $\theta$, and we obtain
\begin{equation}\label{3.35}
\tau_s(\theta,0)=\tan\left(s\frac{\theta}{2}+\frac{\pi}{4}\right).
\end{equation}

The most restrictive is the requirement of invariance under charge conjugation,
\begin{equation}\label{3.36}
C:\,\,\,\,\Phi\rightarrow-\Phi\,\,\,\,\psi\rightarrow {\rm i}\gamma^1\psi^*
\end{equation}
(note that $(\gamma^1)^* = - \gamma^1$ in 
representation \eqref{2.12}), then
\begin{equation}\label{3.37}
K =\tilde K = \pm {\rm i} \gamma^r
\end{equation}
and
\begin{equation}\label{3.38}
|\tau_s| = 1. 
\end{equation}
It should be noted that
\begin{equation}\label{3.39}
K = \tilde K ={\rm i} \gamma^r
\end{equation}
and, consequently,
\begin{equation}\label{3.40}
\tau_s = 1 
\end{equation}
correspond to the quark bag boundary condition, cf. \eqref{3.3}. 

We conclude that the requirement of discrete symmetries reduces the number of boundary parameters. Both $P$ and $CT$ invariances result in the two-parameter position-dependent boundary condition, whereas $CPT$ invariance results in the one-parameter position-dependent boundary condition. The requirement of $C$ invariance is the strongest one, restricting matrix $K$ to the form given by \eqref{3.37}: no position-dependent parameters, but still some ambiguities. In the present study we shall find the requirement restricting matrix $K$ to the unambiguous form given by \eqref{3.39}, i.e. corresponding to the quark bag boundary condition.

\section{Induced vacuum current and magnetic field}\setcounter{equation}{0}

Presenting the angular component of the induced vacuum current as
\begin{equation}\label{4.1}
j_\varphi(r) = \sum_{s = \pm 1} j_\varphi^{(s)}(r),
\end{equation}
where
\begin{equation}\label{4.2}
j_\varphi^{(s)}(r) = - \frac{s \, r}{\nu} \sum\hspace{-1.4em}\int \sum_{n \in
\mathbb{Z}} {\rm sgn} (E) \, f_n^{(s)}(r,E) \, g_n^{(s)}(r,E),
\end{equation}
we note that current \eqref{4.2} can be related to the current in two-dimensional surface $z={\rm const}$, $j_\varphi^{(2{\rm dim})(s)}(r,m)$,
\begin{equation}\label{4.3}
j_\varphi^{(s)}(r) = \int\limits_0^\infty \frac{dk_3}{\pi} \,\, j_\varphi^{(2{\rm dim})(s)}(r,m_3),  \quad  m_3 = \sqrt{m^2+k_3^2} .
\end{equation}
Actually, the latter has been computed at $s=1$ in \cite{SiG}, and the computation at $s=-1$ is analogous with the use of the explicit form of modes of the solution to the Dirac equation, see Appendix A; note that one has to substitute $\tau_{s}$ for 
$\tan\left(\frac{\theta}{2}+\frac{\pi}{4}\right)$ in \cite{SiG}, see \eqref{3.19} and \eqref{3.29}. We present here the result for the case of $\nu \geq 1$ and $|F-\frac12| < \frac{1}{2\nu}$, or $\frac12
\leq \nu <1$ and $|F-\frac12| < 1 - \frac{1}{2\nu}$, 
\begin{multline}\label{4.4}
j_\varphi^{(2{\rm dim})(s)}(r,m)=-\frac{m}{2\pi}\left\{ \frac{1}{2\pi}\int\limits_0^\infty
\frac{du}{\cosh(u/2)}\left[1+\frac1{2mr \cosh(u/2)}\right]{\rm e}^{-2mr
\cosh(u/2)}\right. \\
\times \frac{\cos\left[\nu\left(F-\frac12\right)\pi\right]\sinh(\nu u)
\sinh\left[\nu\left(F-\frac12\right)u\right]+ \sin\left[\nu\left(F-\frac12\right)\pi\right]\sin(\nu \pi)
\cosh\left[\nu\left(F-\frac12\right)u\right]}{\cosh(\nu u)-\cos(\nu
\pi)}\\
- \frac{1}{\nu}\sum_{p=1}^{\left[\!\left| {\nu}/2 \right|\!\right]} \left[1 +\frac{1}{2mr\sin(p\pi/\nu)}\right]\,
\exp[-2mr\sin(p\pi/\nu)] \,\frac{\sin[(2F-1)p\pi]}{\sin(p\pi/\nu)}\\
\Biggl. + \frac{1}{4N} \left(1 +\frac{1}{2mr}\right){\rm e}^{-2mr} 
\sin[(2F-1)N\pi]\, \delta_{\nu, \, 2N}\Biggr\} \\
-\frac{s\,r}{\pi^2}\!\int\limits_m^\infty \!\! \frac{dq\,
q^2}{\sqrt{q^2-m^2}}\left\{\frac12\left[C^{(\wedge)(s)}_{\lambda\left[0, \, s\left(\frac12-F\right)\right]}(qr_0, mr_0) - C^{(\vee)(s)}_{\lambda\left[0, \, s\left(F-\frac12\right)\right]}(qr_0, mr_0)\right] \right.\\
\times K_{\lambda\left(0, \, \frac12-F\right)}(qr)
K_{\lambda\left(0, \, F-\frac12\right)}(qr) \\
+ \sum_{l=1}^{\infty} \left[C^{(\wedge)(s)}_{\lambda\left[l, \, s\left(\frac12-F\right)\right]}(qr_0, mr_0)
K_{\lambda\left[l, \, s\left(\frac12-F\right)\right]} (qr)
K_{\lambda\left[l, \, s\left(\frac12-F\right)\right]-1}(qr) \right. \\
\left.\left. - C^{(\vee)(s)}_{\lambda\left[l, \, s\left(F-\frac12\right)\right]}(qr_0, mr_0) K_{\lambda\left[l, \, s\left(F-\frac12\right)\right]}(qr)K_{\lambda\left[l, \, s\left(F-\frac12\right)\right]-1}(qr)\right]\right\},
\end{multline}
where
\begin{equation}\label{4.44}
\lambda(l, y) = \nu(l + y) + 1/2, 
\end{equation}
\begin{multline}\label{4.5}
C^{(\wedge)(s)}_\rho(v,w)=\left\{v
I_\rho(v)K_\rho(v)\tau_s+w\left[I_\rho(v)K_{\rho-1}(v)-I_{\rho-1}(v)K_\rho(v)\right] \right. \\
\left. - vI_{\rho-1}(v)K_{\rho-1}(v)\tau_s^{-1} \right\} \left[vK^2_\rho(v)\tau_s+2wK_\rho(v)K_{\rho-1}(v) + vK^2_{\rho-1}(v)\tau_s^{-1} \right]^{-1}
\end{multline}
and
\begin{multline}\label{4.6}
C^{(\vee)(s)}_\rho(v,w)=\left\{v
I_\rho(v)K_\rho(v)\tau_s^{-1}+w\left[I_\rho(v)K_{\rho-1}(v)-I_{\rho-1}(v)K_\rho(v)\right]\right.\\
\left. - vI_{\rho-1}(v)K_{\rho-1}(v)\tau_s \right\} \left[v
K^2_\rho(v)\tau_s^{-1}+2wK_\rho(v)K_{\rho-1}(v)+vK^2_{\rho-1}(v)\tau_s
\right]^{-1};
\end{multline}
$I_\rho(u)$ and $K_\rho(u)$ are the modified Bessel functions of order $\rho$ with exponential increase and decrease at large real positive values of their argument, $p$ and $N$ are positive integers, 
$\delta_{\omega, \, \omega'}$ is the Kronecker symbol, $\delta_{\omega, \, \omega'}=0$
at $\omega' \neq \omega$ and $\delta_{\omega, \, \omega} = 1$. Note that relation 
\begin{equation}\label{4.77}
\tau_{-s} = \tau_s^{-1} 
\end{equation}
holds in the case of the $CPT$-invariant boundary condition, 
see \eqref{3.35}, and, as a consequence, we obtain relation
\begin{equation}\label{4.7}
C^{(\vee)(-s)}_\rho(v,w)=C^{(\wedge)(s)}_\rho(v,w)
\end{equation}
in this case.

In general case, inserting \eqref{4.4} into \eqref{4.3}, we perform integration over $k_3$ of terms in the first figure brackets in \eqref{4.4}; as to the remaining terms, we change the order of integration,
$$
\int\limits_0^\infty dk_3 \int\limits_{\sqrt{k_3^2+m^2}}^\infty  dq \,\, = \, \int\limits_m^\infty  dq\,\int\limits_0^{\sqrt{q^2-m^2}} dk_3 \,\, , 
$$
and then introduce variable $\xi = \arcsin \left( k_3 / \sqrt{q^2-m^2}\right)$. As a result, we obtain after summing over $s=\pm 1$:
\begin{multline}\label{4.8}
j_\varphi(r)=-\frac{m^2}{\pi^2}\left\{ \frac{1}{2\pi}\int\limits_0^\infty
\frac{du}{\cosh(u/2)}K_2\left[2mr \cosh(u/2)\right]\right. \\
\times \frac{\cos\left[\nu\left(F-\frac12\right)\pi\right]\sinh(\nu u)
\sinh\left[\nu\left(F-\frac12\right)u\right] + \sin\left[\nu\left(F-\frac12\right)\pi\right]\sin(\nu \pi)
\cosh\left[\nu\left(F-\frac12\right)u\right]}{\cosh(\nu u)-\cos(\nu
\pi)}\\
\left. - \frac{1}{\nu}\sum_{p=1}^{\left[\!\left| {\nu}/2 \right|\!\right]} K_2\left[2mr\sin(p\pi/\nu)\right] \,\frac{\sin[(2F-1)p\pi]}{\sin(p\pi/\nu)}
 + \frac{1}{4N} K_2\left(2mr\right) 
\sin[(2F-1)N\pi]\, \delta_{\nu, \, 2N}\right\} \\
+\frac{2 r}{\pi^3}\!\int\limits_0^{\pi/2} \!\! d{\xi} \!\int\limits_m^\infty \!\! dq\, q^2 
\left\{\frac12\left[C^{(\vee)}_{\lambda\left(0, \, F-\frac12\right)}\left(qr_0, \sqrt{q^2\sin^2\xi+m^2\cos^2\xi}\,\,r_0\right) \right.\right. \\ 
\left. - C^{(\wedge)}_{\lambda \left(0, \, \frac12-F \right)}\left(qr_0, \sqrt{q^2\sin^2\xi+m^2\cos^2\xi}\,\,r_0\right)\right]K_{\lambda\left(0, \, F-\frac12\right)}(qr)
K_{\lambda \left(0, \, \frac12-F \right)}(qr)  \\
+ \sum_{l=1}^{\infty} C^{(\vee)}_{\lambda \left(l, \, F-\frac12 \right)}\left(qr_0, \sqrt{q^2\sin^2\xi+m^2\cos^2\xi}\,\,r_0\right)
K_{\lambda \left(l, \, F-\frac12 \right)} (qr)
K_{\lambda \left(l, \, F-\frac12 \right)-1}(qr) \\
\left. - \sum_{l=1}^{\infty} C^{(\wedge)}_{\lambda \left(l, \, \frac12-F \right)}\left(qr_0, \sqrt{q^2\sin^2\xi+m^2\cos^2\xi}\,\,r_0\right) K_{\lambda \left(l, \, \frac12-F \right)}(qr)K_{\lambda \left(l, \, \frac12-F \right)-1}(qr)\right\},
\end{multline}
where
\begin{equation}\label{4.9}
C^{(\vee)}_\rho(v,w)=\frac12\left[C^{(\vee)(1)}_\rho(v,w)+C^{(\wedge)(-1)}_\rho(v,w)\right]
\end{equation}
and
\begin{equation}\label{4.10}
C^{(\wedge)}_\rho(v,w)=\frac12\left[C^{(\wedge)(1)}_\rho(v,w)+C^{(\vee)(-1)}_\rho(v,w)\right].
\end{equation}
Inserting \eqref{4.8} into \eqref{2.30}, we obtain the magnetic field strength,
\begin{multline}\label{4.11}
B_I(r)=-\frac{e \nu m}{2 \pi^2 r}\left\{ \frac{1}{2\pi}\int\limits_0^\infty
\frac{du}{\cosh^2 (u/2)}K_1\left[2mr \cosh(u/2)\right]\right. \\
\times \frac{\cos\left[\nu\left(F-\frac12\right)\pi\right]\sinh(\nu u)
\sinh\left[\nu\left(F-\frac12\right)u\right] + \sin\left[\nu\left(F-\frac12\right)\pi\right]\sin(\nu \pi)
\cosh\left[\nu\left(F-\frac12\right)u\right]}{\cosh(\nu u)-\cos(\nu
\pi)}\\
\left. - \frac{1}{\nu}\sum_{p=1}^{\left[\!\left| {\nu}/2 \right|\!\right]} K_1\left[2mr\sin(p\pi/\nu)\right] \,\frac{\sin[(2F-1)p\pi]}{\sin^2 (p\pi/\nu)}
 + \frac{1}{4N} K_1\left(2mr\right) 
\sin[(2F-1)N\pi]\, \delta_{\nu, \, 2N}\right\} \\
-\frac{e \nu r}{\pi^3}\!\int\limits_0^{\pi/2} \!\! d{\xi} \!\int\limits_m^\infty \!\! dq\, q^2 \left\{\frac12\left[C^{(\vee)}_{\lambda\left(0, \, F-\frac12\right)}\left(qr_0, \sqrt{q^2\sin^2\xi+m^2\cos^2\xi}\,\,r_0\right) \right.\right. \\ 
\left. - C^{(\wedge)}_{\lambda \left(0, \, \frac12-F \right)}\left(qr_0, \sqrt{q^2\sin^2\xi+m^2\cos^2\xi}\,\,r_0\right)\right] W_{\lambda \left(0, \, \left|F-\frac12\right|\right)}(qr)
\\
+ \sum_{l=1}^{\infty} C^{(\vee)}_{\lambda \left(l, \, F-\frac12 \right)}\left(qr_0, \sqrt{q^2\sin^2\xi+m^2\cos^2\xi}\,\,r_0\right)
W_{\lambda \left(l, \, F-\frac12 \right)} (qr)
\\
\left. - \sum_{l=1}^{\infty} C^{(\wedge)}_{\lambda \left(l, \, \frac12-F \right)}\left(qr_0, \sqrt{q^2\sin^2\xi+m^2\cos^2\xi}\,\,r_0\right) W_{\lambda \left(l, \, \frac12-F \right)}(qr)\right\},
\end{multline}
where
\begin{equation}\label{4.112}
W_\rho (v) = K_{\rho}(v)\frac{d}{d \rho} K_{\rho-1}(v) -
K_{\rho-1}(v)\frac{d}{d \rho} K_{\rho}(v).
\end{equation}
Note that both the current and the magnetic field strength change sign under simultaneous change $F \rightarrow 1 - F$ and $\tau_s \rightarrow \tau_s^{-1}$.

The cases when there are no peculiar modes are considered similarly. In the case of $\nu>1$ and $0 < F <
\frac12\left(1-\frac1\nu\right)$, we get
\begin{multline}\label{4.12}
j_\varphi(r)= - \frac{m^2}{\pi^2}\left\{ \frac{1}{2\pi}\int\limits_0^\infty
\frac{du}{\cosh(u/2)}K_2\left[2mr \cosh(u/2)\right]\right. \\
\times \frac{\cos\left[\nu\left(F-\frac12\right)\pi\right] \cosh\left[\nu\left(F+\frac12 \right)u\right]
- \cos[\nu\left(F+\frac12 \right)\pi)]
\cosh\left[\nu\left(F-\frac12\right)u\right]}{\cosh(\nu
u)-\cos(\nu \pi)}\\
\left. - \frac{1}{\nu}\sum_{p=1}^{\left[\!\left| {\nu}/2 \right|\!\right]} K_2\left[2mr\sin(p\pi/\nu)\right] \,\frac{\sin[(2F-1)p\pi]}{\sin(p\pi/\nu)}
 + \frac{1}{4N} K_2\left(2mr\right) 
\sin[(2F-1)N\pi]\, \delta_{\nu, \, 2N}\right\} \\
+\frac{2 r}{\pi^3}\!\int\limits_0^{\pi/2} \!\! d{\xi} \!\int\limits_m^\infty \!\! dq\, q^2\\
\times \left[ \sum_{l=1}^{\infty} C^{(\vee)}_{\lambda \left(l, \, F-\frac12 \right)}\left(qr_0, \sqrt{q^2\sin^2\xi+m^2\cos^2\xi}\,\,r_0\right)
K_{\lambda \left(l, \, F-\frac12 \right)} (qr)
K_{\lambda \left(l, \, F-\frac12 \right)-1}(qr) \right. \\
\left. - \sum_{l=0}^{\infty} C^{(\wedge)}_{\lambda \left(l, \, \frac12-F \right)}\left(qr_0, \sqrt{q^2\sin^2\xi+m^2\cos^2\xi}\,\,r_0\right) K_{\lambda \left(l, \, \frac12-F \right)}(qr)K_{\lambda \left(l, \, \frac12-F \right)-1}(qr)\right]
\end{multline}
and
\begin{multline}\label{4.13}
B_I (r)=-\frac{e \nu m}{2 \pi^2 r}\left\{ \frac{1}{2\pi}\int\limits_0^\infty
\frac{du}{\cosh^2(u/2)} K_1\left[2mr \cosh(u/2)\right]\right. \\
\times \frac{\cos\left[\nu\left(F-\frac12\right)\pi\right] \cosh\left[\nu\left(F+\frac12 \right)u\right]
- \cos[\nu\left(F+\frac12 \right)\pi)]
\cosh\left[\nu\left(F-\frac12\right)u\right]}{\cosh(\nu
u)-\cos(\nu \pi)}\\
\left. - \frac{1}{\nu}\sum_{p=1}^{\left[\!\left| {\nu}/2 \right|\!\right]} K_1\left[2mr\sin(p\pi/\nu)\right] \,\frac{\sin[(2F-1)p\pi]}{\sin^2(p\pi/\nu)}
 + \frac{1}{4N} K_1\left(2mr\right) 
\sin[(2F-1)N\pi]\, \delta_{\nu, \, 2N}\right\} \\
-\frac{e \nu r}{\pi^3}\!\int\limits_0^{\pi/2} \!\! d{\xi} \!\int\limits_m^\infty \!\! dq\,q^2 \left[ \sum_{l=1}^{\infty} C^{(\vee)}_{\lambda \left(l, \, F-\frac12 \right)}\left(qr_0, \sqrt{q^2\sin^2\xi+m^2\cos^2\xi}\,\,r_0\right)
W_{\lambda \left(l, \, F-\frac12 \right)} (qr)
 \right. \\
\left. - \sum_{l=0}^{\infty} C^{(\wedge)}_{\lambda \left(l, \, \frac12-F \right)}\left(qr_0, \sqrt{q^2\sin^2\xi+m^2\cos^2\xi}\,\,r_0\right) W_{\lambda \left(l, \, \frac12-F \right)}(qr)\right].
\end{multline}
In the case of $\nu>1$ and $\frac12\left(1+\frac1\nu \right) < F < 1$, we get
\begin{multline}\label{4.14}
j_\varphi(r)=\frac{m^2}{\pi^2}\left\{ \frac{1}{2\pi}\int\limits_0^\infty
\frac{du}{\cosh(u/2)}K_2\left[2mr \cosh(u/2)\right]\right. \\
\times \frac{\cos\left[\nu\left(F-\frac12\right)\pi\right]
\cosh\left[\nu\left(F-\frac32 \right)u\right] -
\cos[\nu\left(F-\frac32 \right)\pi)]
\cosh\left[\nu\left(F-\frac12\right)u\right]}{\cosh(\nu u)-\cos(\nu
\pi)}\\
\left. + \frac{1}{\nu}\sum_{p=1}^{\left[\!\left| {\nu}/2 \right|\!\right]} K_2\left[2mr\sin(p\pi/\nu)\right] \,\frac{\sin[(2F-1)p\pi]}{\sin(p\pi/\nu)}
 - \frac{1}{4N} K_2\left(2mr\right) 
\sin[(2F-1)N\pi]\, \delta_{\nu, \, 2N}\right\} \\
+\frac{2r}{\pi^3}\!\int\limits_0^{\pi/2} \!\! d{\xi} \!\int\limits_m^\infty \!\! dq\, q^2 \\
\times \left[ \sum_{l=0}^{\infty} C^{(\vee)}_{\lambda \left(l, \, F-\frac12 \right)}\left(qr_0, \sqrt{q^2\sin^2\xi+m^2\cos^2\xi}\,\,r_0\right)
K_{\lambda \left(l, \, F-\frac12 \right)} (qr)
K_{\lambda \left(l, \, F-\frac12 \right)-1}(qr) \right. \\
\left. - \sum_{l=1}^{\infty} C^{(\wedge)}_
{\lambda \left(l, \, \frac12-F \right)}\left(qr_0, \sqrt{q^2\sin^2\xi+m^2\cos^2\xi}\,\,r_0\right) K_{\lambda \left(l, \, \frac12-F \right)}(qr)K_{\lambda \left(l, \, \frac12-F \right)-1}(qr)\right]
\end{multline}
and
\newpage
\begin{multline}\label{4.15}
B_I (r)=\frac{e \nu m}{2 \pi^2 r}\left\{ \frac{1}{2\pi}\int\limits_0^\infty
\frac{du}{\cosh^2(u/2)}K_1\left[2mr \cosh(u/2)\right]\right. \\
\times \frac{\cos\left[\nu\left(F-\frac12\right)\pi\right]
\cosh\left[\nu\left(F-\frac32 \right)u\right] -
\cos[\nu\left(F-\frac32 \right)\pi)]
\cosh\left[\nu\left(F-\frac12\right)u\right]}{\cosh(\nu u)-\cos(\nu
\pi)}\\
\left. + \frac{1}{\nu}\sum_{p=1}^{\left[\!\left| {\nu}/2 \right|\!\right]} K_1\left[2mr\sin(p\pi/\nu)\right] \,\frac{\sin[(2F-1)p\pi]}{\sin^2(p\pi/\nu)}
 - \frac{1}{4N} K_1\left(2mr\right) 
\sin[(2F-1)N\pi]\, \delta_{\nu, \, 2N}\right\} \\
-\frac{e \nu r}{\pi^3}\!\int\limits_0^{\pi/2} \!\! d{\xi} \!\int\limits_m^\infty \!\! dq\, q^2  \left[ \sum_{l=0}^{\infty} C^{(\vee)}_{\lambda \left(l, \, F-\frac12 \right)}\left(qr_0, \sqrt{q^2\sin^2\xi+m^2\cos^2\xi}\,\,r_0\right)
W_{\lambda \left(l, \, F-\frac12 \right)} (qr)
 \right. \\
\left. - \sum_{l=1}^{\infty} C^{(\wedge)}_{\lambda \left(l, \, \frac12-F \right)}\left(qr_0, \sqrt{q^2\sin^2\xi+m^2\cos^2\xi}\,\,r_0\right) W_{\lambda \left(l, \, \frac12-F \right)}(qr)\right].
\end{multline}
Note that \eqref{4.12} and \eqref{4.13} turn into \eqref{4.14} and \eqref{4.15} with opposite sign under simultaneous change $F \rightarrow 1 - F$ and $\tau_s \rightarrow \tau_s^{-1}$.

In the case of the vanishing string tension, the expressions for the current and the magnetic field strength are simplified:
\begin{multline}\label{4.16}
\left.j_\varphi(r)\right|_{\nu=1} = -\frac{m^2}{\pi^3}\sin(F\pi)\int\limits_1^\infty dv \, v^{-2} \, K_2\left(2mrv\right) \sinh\left[\left(2F - 1\right){\rm arccosh} 
v \right]\\
+\frac{2 r}{\pi^3}\!\int\limits_0^{\pi/2} \!\! d{\xi} \!\int\limits_m^\infty \!\! dq\, q^2 \left\{ \frac12 \left[C^{(\vee)}_{F}\left(qr_0, \sqrt{q^2\sin^2\xi+m^2\cos^2\xi}\,\,r_0\right) \right. \right. \\
\left. - C^{(\wedge)}_{1-F}\left(qr_0, \sqrt{q^2\sin^2\xi+m^2\cos^2\xi}\,\,r_0\right)\right] K_{F}(qr) K_{1 - F}(qr) \\
+ \sum_{l=1}^{\infty} C^{(\vee)}_{l+F}\left(qr_0, \sqrt{q^2\sin^2\xi+m^2\cos^2\xi}\,\,r_0\right)
K_{l+F} (qr) K_{l - 1 + F}(qr)  \\
\left. - \sum_{l=1}^{\infty}C^{(\wedge)}_{l+1-F}\left(qr_0, \sqrt{q^2\sin^2\xi+m^2\cos^2\xi}\,\,r_0\right) K_{l+1-F}(qr)K_{l-F}(qr)\right\}
\end{multline}
and
\newpage
\begin{multline}\label{4.17}
\left.B_I(r)\right|_{\nu=1} = -\frac{e \nu m}{2 \pi^3 r}\sin(F\pi)\int\limits_1^\infty dv \, v^{-3} \, K_1\left(2mrv\right) \sinh\left[\left(2F - 1\right){\rm arccosh} 
v \right]\\
-\frac{e \nu r}{\pi^3}\!\int\limits_0^{\pi/2} \!\! d{\xi} \!\int\limits_m^\infty \!\! dq\, q^2\left\{ \frac12 \left[C^{(\vee)}_{F}\left(qr_0, \sqrt{q^2\sin^2\xi+m^2\cos^2\xi}\,\,r_0\right) \right. \right. \\
\left. - C^{(\wedge)}_{1-F}\left(qr_0, \sqrt{q^2\sin^2\xi+m^2\cos^2\xi}\,\,r_0\right)\right] W_{\frac12+\left|F-\frac12\right|} (qr)  \\
+ \sum_{l=1}^{\infty} C^{(\vee)}_{l+F}\left(qr_0, \sqrt{q^2\sin^2\xi+m^2\cos^2\xi}\,\,r_0\right)
W_{l+F} (qr)  \\
\left. - \sum_{l=1}^{\infty} C^{(\wedge)}_{l+1-F}\left(qr_0, \sqrt{q^2\sin^2\xi+m^2\cos^2\xi}\,\,r_0\right) W_{l+1-F}(qr)\right\}.
\end{multline}

In Appendix B the above results at $\nu \geq 1$ and $|F-\frac12| < \frac{1}{2\nu}$, or at $\frac12
\leq \nu <1$ and $|F-\frac12| < 1 - \frac{1}{2\nu}$, are presented for the cases $F \neq 1/2$ and $F = 1/2$ separately. This allows us to collect the dependence on the transverse size of a cosmic string in pieces denoted by $j_\varphi^{(b)(s)}$ and $B_I^{(b)(s)}$: 
\begin{equation}\label{4.18}
j_\varphi^{(s)}(r)= j_\varphi^{(a)(s)}(r) + j_\varphi^{(b)(s)}(r),\qquad
j_\varphi^{(a)(s)}(r)=\lim_{r_0\rightarrow 0}j_\varphi^{(s)}(r)
\end{equation}
and
\begin{equation}\label{4.19}
B_I^{(s)}(r)= B_I^{(a)(s)}(r) + B_I^{(b)(s)}(r),\qquad
B_I^{(a)(s)}(r)=\lim_{r_0\rightarrow 0}B_I^{(s)}(r);
\end{equation}
then it is clear that $j_\varphi^{(a)(s)}(s)$ and $B_I^{(a)(s)}(r)$ correspond to the case of the infinitely thin cosmic string, which is obtained by imposing the condition of minimal irregularity on peculiar mode \eqref{a5} in the $r_0 \rightarrow 0$ limit, see Appendix A; then the mode coefficient takes the form of \eqref{c3}. The current and the magnetic field strength in the case of the infinitely thin cosmic string are given in Appendix C. 
Returning to the string with nonvanishing transverse size, we note that the same structute, as that of \eqref{4.18} and \eqref{4.19}, is evident at $\nu > 1$ and $\frac{1}{2\nu} <|F-\frac12|< \frac{1}{2}$, see \eqref{4.12}-\eqref{4.15}. In the case of $F \neq 1/2$, the $r_0$-independent pieces are independent of $\tau_s$, see \eqref{c4} and \eqref{c5}, whereas the $r_0$-dependent pieces, consisting of the integrals over $q$ in \eqref{b1}, \eqref{b2}, \eqref{b11}, \eqref{b12}, and \eqref{4.12}-\eqref{4.15}, depend on $\tau_s$. In the case of $F = 1/2$, the $r_0$-independent pieces depend on $\tau_s$, see \eqref{c9} and \eqref{c10}, while the $r_0$-dependent pieces, consisting of the integrals over $q$ and the appropriate parts of the other terms in \eqref{b7} and \eqref{b13}, depend on $\tau_s$ as well. Let us recall that boundary parameter $\tau_s$ in general depends on $\varphi$ and $z$, see Section 3. 

As a consequence of the above, limits $F\rightarrow 1/2$ and $r_0\rightarrow 0$ in most cases do not commute. If limit $r_0\rightarrow 0$ is taken first, we get a discontinuity at $F=1/2$,
\begin{equation}\label{4.20}
\left.\lim_{F\rightarrow (1/2)_{\pm}}
\lim_{r_0\rightarrow 0}j_\varphi^{(s)}(r)\right|_{F \neq 1/2, \, \ln|\tau_s|^s \neq \pm \infty}\!=\pm \frac{1}{(4\pi)^{2}r^2} \left(1+2mr\right){\rm e}^{-2mr}
\end{equation}
and
\begin{equation}\label{4.21}
\left.\lim_{F\rightarrow (1/2)_{\pm}}
\lim_{r_0\rightarrow 0}B_I^{(s)}(r)\right|_{F \neq 1/2, \, \ln|\tau_s|^s \neq \pm \infty}\!= \pm \frac{e \nu}{2 (4\pi)^{2}r^2} \left[\left(1+2mr\right){\rm e}^{-2mr} - (2mr)^2 \Gamma(0,2mr)\right],
\end{equation}
where 
$$\Gamma(t,u)=\int\limits_u^\infty dy\,y^{t-1}{\rm e}^{-y} $$
is the incomplete gamma-function. If the order of limits is reversed, then we get
\begin{equation}\label{4.22}
\lim_{r_0\rightarrow0}\, \left.\lim_{F\rightarrow (1/2)_{\pm}}
j_\varphi^{(s)}(r)\right|_{F \neq 1/2, \, \ln|\tau_s|^s \neq \pm \infty}\,= \lim_{r_0\rightarrow0}\,\left.
j_\varphi^{(s)}(r)\right|_{F=1/2,\, \ln|\tau_s|^s \neq \pm \infty}
\end{equation}
and
\begin{equation}\label{4.23}
\lim_{r_0\rightarrow0}\, \left.\lim_{F\rightarrow (1/2)_{\pm}}
B_I^{(s)}(r)\right|_{F \neq 1/2, \, \ln|\tau_s|^s \neq \pm \infty}\!= \lim_{r_0\rightarrow0}\,\left.
B_I^{(s)}(r)\right|_{F=1/2,\, \ln|\tau_s|^s \neq \pm \infty},
\end{equation}
that are depending on $\tau_s$, cf. \eqref{c9} and \eqref{c10}. The limits do commute in special cases only:
\begin{equation}\label{4.24}
\left.\lim_{F\rightarrow (1/2)_{\pm}}
\lim_{r_0\rightarrow 0}j_\varphi^{(s)}(r)\right|_{\ln|\tau_s|^s = \pm \infty}\!=
\left.\lim_{r_0\rightarrow 0} \, \lim_{F\rightarrow (1/2)_{\pm}}
j_\varphi^{(s)}(r)\right|_{\ln|\tau_s|^s = \mp \infty}\!=\mp \frac{1}{(4\pi)^{2}r^2} \left(1+2mr\right){\rm e}^{-2mr}
\end{equation}
and
\begin{multline}\label{4.25}
\left.\lim_{F\rightarrow (1/2)_{\pm}}
\lim_{r_0\rightarrow 0}B_I^{(s)}(r)\right|_{\ln|\tau_s|^s = \pm \infty}\!=
\left.\lim_{r_0\rightarrow 0} \, \lim_{F\rightarrow (1/2)_{\pm}}
B_I^{(s)}(r)\right|_{\ln|\tau_s|^s = \pm \infty}\\
= \mp \frac{e \nu}{2 (4\pi)^{2}r^2} \left[\left(1+2mr\right){\rm e}^{-2mr} - (2mr)^2 \Gamma(0,2mr)\right];
\end{multline}
the discontinuity at $F=1/2$ is absent in these cases.

The temporal component of the induced vacuum current is considered in a similar way. Presenting \eqref{2.27} as
\begin{equation}\label{4.26}
j^0 (r) = \sum_{s = \pm 1} j^{0 (s)} (r),
\end{equation}
where
\begin{equation}\label{4.27}
j^{0 (s)}(r) = - \frac 12 \sum\hspace{-1.4em}\int \sum_{n \in
\mathbb{Z}} {\rm sgn} (E) \left\{\left[f_n^{(s)}(r,E)\right]^2 + \left[g_n^{(s)}(r,E)\right]^2\right\},
\end{equation}
we obtain in the case of $\nu \geq 1$ and $|F-\frac12| < \frac{1}{2\nu}$, or $\frac12
\leq \nu <1$ and $|F-\frac12| < 1 - \frac{1}{2\nu}$:
\begin{multline}\label{4.28}
j^{0(s)}(r)= - \frac{s \nu m^2}{2 \pi^2 r}
\left\{\frac{1}{2\pi}\int\limits_0^\infty du \left[K_0\Biggl(2mr \cosh(u/2)\Biggr) 
+ \frac{1}{2mr \cosh(u/2)}K_1\Biggl(2mr \cosh(u/2)\Biggr)\right] \right. \\ 
\times \Biggl[\frac{\cos\left[\nu\left(F-\frac12\right)\pi\right]\sinh(\nu u)
\sinh\left[\nu\left(F-\frac12\right)u\right] + \sin\left[\nu\left(F-\frac12\right)\pi\right]\sin(\nu \pi)
\cosh\left[\nu\left(F-\frac12\right)u\right]}{\cosh(\nu u)-\cos(\nu
\pi)} \Biggr. \\
\Biggl. - \cos\left[\nu\left(F-\frac12\right)\pi\right]\tanh\left(\frac{u}{2}\right)\sinh\left[\nu\left(F-\frac12\right)u\right] \Biggr] \\
- \frac{1}{\nu}\sum_{p=1}^{\left[\!\left| {\nu}/2 \right|\!\right]} \left[K_0\Biggl(2mr\sin(p\pi/\nu)\Biggr) + \frac{1}{2mr \sin(p\pi/\nu)}K_1\Biggl(2mr\sin(p\pi/\nu)\Biggr)\right] \,\sin[(2F-1)p\pi] \\
\left.  + \frac{1}{4N} \left[K_0\left(2mr\right) + \frac{1}{2mr}K_1\left(2mr\right)\right] 
\sin[(2F-1)N\pi]\, \delta_{\nu, \, 2N}\right\} \\
+\frac{\nu}{2 \pi^3}\!\int\limits_0^{\pi/2} \!\! d{\xi} \!\int\limits_m^\infty \!\! dq\, q \,  
\sqrt{q^2\sin^2\xi+m^2\cos^2\xi} \\
\times \left\{C^{(\vee)(s)}_{\lambda\left[0, \, s\left(F-\frac12\right)\right]}\left(qr_0, \sqrt{q^2\sin^2\xi+m^2\cos^2\xi}\,r_0\right)K^2_{\lambda\left[0, \, s\left(F-\frac12\right)\right]}(qr) \right. \\ 
- C^{(\wedge)(s)}_{\lambda\left[0, \, s\left(\frac12-F\right)\right]}\left(qr_0, \sqrt{q^2\sin^2\xi+m^2\cos^2\xi}\,r_0\right)K^2_{\lambda\left[0, \, s\left(\frac12-F\right)\right]}(qr)  \\
 + \sum_{l=1}^{\infty} C^{(\vee)(s)}_{\lambda\left[l, \, s\left(F-\frac12\right)\right]}\left(qr_0, \sqrt{q^2\sin^2\xi+m^2\cos^2\xi}\,r_0\right)\left[K^2_{\lambda\left[l, \, s\left(F-\frac12\right)\right]}(qr) - K^2_{\lambda\left[l, \, s\left(F-\frac12\right)\right]-1}(qr)\right] \\
 \left. - \sum_{l=1}^{\infty} C^{(\wedge)(s)}_{\lambda\left[l, \, s\left(\frac12-F\right)\right]}\left(qr_0, \sqrt{q^2\sin^2\xi+m^2\cos^2\xi}\,r_0\right)\left[K^2_{\lambda\left[l, \, s\left(\frac12-F\right)\right]} (qr) - K^2_{\lambda\left[l, \, s\left(\frac12-F\right)\right]-1}(qr)\right]\right\}.
\end{multline}
Summing over $s$, we get
\begin{multline}\label{4.29}
j^0 (r)= \frac{\nu}{\pi^3}\!\int\limits_0^{\pi/2} \!\! d{\xi} \!\int\limits_m^\infty \!\! dq\, q   \sqrt{q^2\sin^2\xi+m^2\cos^2\xi} \left\{{\tilde C}^{(\vee)}_{\lambda\left(0, \, F-\frac12\right)}\left(qr_0, \sqrt{q^2\sin^2\xi+m^2\cos^2\xi}\,r_0\right)\right. \\ 
\times K^2_{\lambda\left(0, \, F-\frac12\right)}(qr) - {\tilde C}^{(\wedge)}_{\lambda \left(0, \, \frac12-F \right)}\left(qr_0, \sqrt{q^2\sin^2\xi+m^2\cos^2\xi}\,r_0\right)
K^2_{\lambda \left(0, \, \frac12-F \right)}(qr) \\
+ \sum_{l=1}^{\infty} {\tilde C}^{(\vee)}_{\lambda \left(l, \, F-\frac12 \right)}\left(qr_0, \sqrt{q^2\sin^2\xi+m^2\cos^2\xi}\,r_0\right) \left[K^2_{\lambda \left(l, \, F-\frac12 \right)} (qr) - K^2_{\lambda \left(l, \, F-\frac12 \right)-1}(qr) \right] \\
\left. - \sum_{l=1}^{\infty}{\tilde C}^{(\wedge)}_{\lambda \left(l, \, \frac12-F \right)}\left(qr_0, \sqrt{q^2\sin^2\xi+m^2\cos^2\xi}\,r_0\right) \left[K^2_{\lambda \left(l, \, \frac12-F \right)}(qr) - K^2_{\lambda \left(l, \, \frac12-F \right)-1}(qr)\right]\right\},
\end{multline}
where
\begin{equation}\label{4.30}
{\tilde C}^{(\vee)}_\rho(v,w)=\frac12\left[C^{(\vee)(1)}_\rho(v,w) - C^{(\wedge)(-1)}_\rho(v,w)\right]
\end{equation}
and
\begin{equation}\label{4.31}
{\tilde C}^{(\wedge)}_\rho(v,w)=\frac12\left[C^{(\wedge)(1)}_\rho(v,w) - C^{(\vee)(-1)}_\rho(v,w)\right].
\end{equation}
Also, we obtain in the case of $\nu>1$ and $0 < F <
\frac12\left(1-\frac1\nu\right)$
\begin{multline}\label{4.32}
j^0 (r)= \frac{\nu}{\pi^3}\!\int\limits_0^{\pi/2} \!\! d{\xi} \!\int\limits_m^\infty \!\! dq\, q   \sqrt{q^2\sin^2\xi+m^2\cos^2\xi} \\ 
\times \left\{\sum_{l=1}^{\infty} {\tilde C}^{(\vee)}_{\lambda \left(l, \, F-\frac12 \right)}\left(qr_0, \sqrt{q^2\sin^2\xi+m^2\cos^2\xi}\,r_0\right) \left[K^2_{\lambda \left(l, \, F-\frac12 \right)} (qr) - K^2_{\lambda \left(l, \, F-\frac12 \right)-1}(qr) \right] \right. \\
\left. - \sum_{l=0}^{\infty}{\tilde C}^{(\wedge)}_{\lambda \left(l, \, \frac12-F \right)}\left(qr_0, \sqrt{q^2\sin^2\xi+m^2\cos^2\xi}\,r_0\right) \left[K^2_{\lambda \left(l, \, \frac12-F \right)}(qr) - K^2_{\lambda \left(l, \, \frac12-F \right)-1}(qr)\right]\right\}
\end{multline}
and in the case of $\nu>1$ and $\frac12\left(1+\frac1\nu \right) < F < 1$
\begin{multline}\label{4.33}
j^0 (r)= \frac{\nu}{\pi^3}\!\int\limits_0^{\pi/2} \!\! d{\xi} \!\int\limits_m^\infty \!\! dq\, q   \sqrt{q^2\sin^2\xi+m^2\cos^2\xi} \\
\times \left\{\sum_{l=0}^{\infty} {\tilde C}^{(\vee)}_{\lambda \left(l, \, F-\frac12 \right)}\left(qr_0, \sqrt{q^2\sin^2\xi+m^2\cos^2\xi}\,r_0\right) \left[K^2_{\lambda \left(l, \, F-\frac12 \right)} (qr) - K^2_{\lambda \left(l, \, F-\frac12 \right)-1}(qr) \right] \right. \\
\left. - \sum_{l=1}^{\infty}{\tilde C}^{(\wedge)}_{\lambda \left(l, \, \frac12-F \right)}\left(qr_0, \sqrt{q^2\sin^2\xi+m^2\cos^2\xi}\,r_0\right) \left[K^2_{\lambda \left(l, \, \frac12-F \right)}(qr) - K^2_{\lambda \left(l, \, \frac12-F \right)-1}(qr)\right]\right\}.
\end{multline}
Note that coefficients ${\tilde C}^{(\vee)}_\rho(v,w)$ \eqref{4.30} and ${\tilde C}^{(\wedge)}_\rho(v,w)$ \eqref{4.31} depend on boundary parameters $\tau_1$ and 
$\tau_{-1}$, the latters in general are varying with $\varphi$ and $z$. Clearly, any variation of $\tau_s$ with $z$ cannot ensure a sufficient decrease (as power $|z|^{- 1 - \varepsilon}, \varepsilon > 0$) of 
$j^0 (r)$ at $z \rightarrow \pm \infty$ in order to make total charge $Q_{\rm I}$ \eqref{2.29} to be finite. Therefore, as long as ${\tilde C}^{(\vee)}_\rho(v,w)$ and ${\tilde C}^{(\wedge)}_\rho(v,w)$ are nonvanishing, $Q_{\rm I}$ is infinite. However, 
${\tilde C}^{(\vee)}_\rho(v,w)$ and ${\tilde C}^{(\wedge)}_\rho(v,w)$ vanish in the case of the $CPT$ invariant boundary condition, as a consequence of \eqref{4.7}; therefore, $j^0 (r)$ and $Q_{\rm I}$ are exactly zero in this case.

In general case, the temporal component of the current at $F \neq 1/2$ is zero in the limit of the vanishing transverse size of a cosmic string; the expression at $F = 1/2$ in this case is given in Appendix C, see \eqref{c17}; note that $\left. \nu ^{-1} j^{0}(r)\right|_{r_0 = 0, \, F=1/2}$ is independent of the string tension.

\section{Finiteness of the induced vacuum magnetic flux}\setcounter{equation}{0}

The induced vacuum current and the induced vacuum magnetic field strength decrease exponentially at large distances from a cosmic string. So the crucial point for the finiteness of the total flux of the induced vacuum magnetic field is the behaviour of the local characteristics in the vicinity of the string location. 

In the case of the infinitely thin cosmic string, the current and, consequently, the magnetic field strength increase in the vicinity as the inverse squared distance from the string location [see \eqref{c3}, \eqref{c4}, \eqref{c8}, and \eqref{c9} in Appendix C]; as a result, the flux is infinite. If a transverse size of a string is introduced, this opens an opportunity to make  the flux finite, but the opportunity by no means is realized for all boundary conditions. The requirement of the flux finiteness can be formulated as 
\begin{equation}\label{5.1}
\lim_{r \rightarrow r_0}j_\varphi(r) \, (r-r_0)^2 = 0,
\end{equation}
or
\begin{equation}\label{5.2}
\lim_{r \rightarrow r_0}B_I(r) \, (r-r_0) = 0,
\end{equation}
then, using integration by parts, flux \eqref{2.31} can be presented as
\begin{equation}\label{5.3}
\Phi_{\rm I} = \frac{e}{2} \int\limits_{0}^{2\pi} d \varphi \int\limits_{r_0}^\infty \frac{dr}{r} \,
j_\varphi(r) \, (r^2-r_0^2).
\end{equation}
The $r \rightarrow r_0$ asymptotics of the current in two-dimensional surface $z = {\rm const}$ was studied in \cite{SiG}, and, as follows from this study, condition
\begin{equation}\label{5.4}
\lim_{r \rightarrow r_0}j^{(2{\rm dim})(s)}_\varphi(r, m) \, (r-r_0)^2 = 0
\end{equation}
is maintained at $|\tau_s| = 1$ only. Integration of $j^{(2{\rm dim})(s)}_\varphi(r, \sqrt {m^2 + k^2_3})$ over $k_3$, see \eqref{4.3}, is hardly to yield zero, however it can yield divergence at $r \rightarrow r_0$. Thus, we have no doubt that conditions \eqref{5.1} and \eqref{5.2} are invalid at $|\tau_s| \neq 1$ and have to check their validity at $|\tau_s| = 1$. As is shown in Appendix B, expression \eqref{b6} that is finite at $r \rightarrow r_0$ turns, after integration, into expression \eqref{b10} that is quadratically diveregent at $r \rightarrow r_0$; consequently,
expression \eqref{b15} is linearly diveregent at $r \rightarrow r_0$. We conclude that, in the case of $F= 1/2$, conditions \eqref{5.1} and \eqref{5.2} are violated and the flux is infinite at $\tau_s \neq 1$. Otherwise, we obtain, see \eqref{b9} and \eqref{b14},
\begin{equation}\label{5.5}
\left.\Phi_{\rm
I}\right|_{F=1/2,\,\tau_1=\tau_{-1}=1} = 0.
\end{equation}

In the case of $F \neq 1/2$, conditions \eqref{5.1} and \eqref{5.2} are maintained at $|\tau_s| = 1$. Defining 
\begin{multline}\label{5.6}
C^{(\pm)}_\rho(v, w)=\left\{v
I_\rho(v)K_\rho(v) \pm w\left[I_\rho(v)K_{\rho-1}(v)-I_{\rho-1}(v)K_\rho(v)\right] -
vI_{\rho-1}(v)K_{\rho-1}(v)
\right\} \\
\times \left[v
K^2_\rho(v)\pm 2wK_\rho(v)K_{\rho-1}(v) + vK^2_{\rho-1}(v)\right]^{-1}
\end{multline}
and performing integration over 
$\varphi$ and $r$ (see \cite{SiG} for details), we obtain
\begin{multline}\label{5.7}
\left. \Phi_{\rm I}\right|_{F \neq 1/2, \, \tau_1=\tau_{-1} = \pm 1} = \frac{e}{2 \pi}\left\{\frac{1}{2\pi} \int\limits_0^\infty
\frac{du}{\cosh^3(u/2)}\,K_0[2mr_0 \cosh(u/2)]\, \Omega_{{\rm sgn}\left(F - \frac12\right)}(u)\right. \\
\left.+ \frac{1}{\nu}\sum_{p=1}^{\left[\!\left| {\nu}/2
\right|\!\right]} K_0[2mr_0\sin(p\pi/\nu)]
\,\frac{\sin[(2F-1)p\pi]}{\sin^3(p\pi/\nu)} - \frac{1}{4N} K_0 (2mr_0)
\sin\left[\left(2F -1\right) N \pi \right] \,
\delta_{\nu, \, 2N}\right\} \\ 
+ \frac{e}{\pi^2} \!\int\limits_{0}^{\pi/2} \!d\xi\!\int\limits_{m
r_0}^{\infty}\!dv\, v \Biggl\{{\rm sgn}\!\left(F-\frac12\right)C^{(\pm)}_{\lambda \left(0, \, \left|F-\frac12\right| \right)}\left(v, \sqrt{v^2\sin^2\xi+m^2 r_0^2\cos^2\xi}\right)
D_{\lambda \left(0, \, \left|F-\frac12\right|\right)}(v) \Biggr. \\
+ \sum_{l=1}^\infty \Biggl[C^{(\pm)}_{\lambda \left(l, \, F-\frac12 \right)}\left(v, \sqrt{v^2\sin^2\xi+m^2 r_0^2\cos^2\xi}\right) D_{\lambda \left(l, \, F-\frac12 \right)}(v) \Biggr. \\
\Biggl. \Biggl.  - C^{(\pm)}_{\lambda \left(l, \, \frac12-F \right)}\left(v, \sqrt{v^2\sin^2\xi+m^2 r_0^2\cos^2\xi}\right)
D_{\lambda \left(l, \, \frac12-F \right)}(v) \Biggr]\Biggr\} 
\end{multline}
and
\begin{equation}\label{5.8}
\left. \Phi_{\rm I}\right|_{F \neq 1/2, \, \tau_1= -\tau_{-1} = \pm 1} = \frac12\left(\left. \Phi_{\rm I}\right|_{F \neq 1/2, \, \tau_1=\tau_{-1} = 1} + \left. \Phi_{\rm I}\right|_{F \neq 1/2, \, \tau_1=\tau_{-1} = -1}\right), 
\end{equation}
where 
\begin{equation}\label{5.9}
D_{\rho}(v) = \rho K_\rho^2(v)-(\rho-1)K_{\rho+1}(v)K_{\rho-1}(v)
+ v W_{\rho}(v),
\end{equation}
$W_{\rho}(v)$ is given by \eqref{4.112}, and
\begin{multline}\label{5.10}
\Omega_{\pm 1}(u) \\
=\pm \frac{\cos\left[\nu\left(F-\frac12\right)\pi\right] \cosh\left[\nu\left(F-\frac12 \mp 1\right)u\right]
- \cos\left[\nu\left(F-\frac12 \mp 1\right)\pi\right]\cosh\left[\nu\left(F-\frac12\right)u\right]}{\cosh(\nu u)-\cos(\nu \pi)}.
\end{multline}
Note that the flux changes sign under $F \rightarrow 1-F$; as a consequence, it has to vanish at $F = 1/2$, that is confirmed at 
$\tau_1=\tau_{-1} = 1$ by direct calculation, see \eqref{5.5}. The absolute value of the flux at $F \neq 1/2$ increases with the increase of $\nu$, with $e^{-1} \Phi_I$ being positive at $F > 1/2$ and negative at $F < 1/2$.

In the case of the vanishing string tension, the expression for the flux is simplified,
\begin{multline}\label{5.11}
\left. \Phi_{\rm I}\right|_{\nu = 1, \, F \neq 1/2, \, \tau_1=\tau_{-1} = \pm 1} = \frac{e}{2 \pi^2} \, {\rm sgn}\left(F - \frac12\right) \, \sin(F\pi) \int\limits_1^\infty
\frac{dv}{v^4 \sqrt{v^2-1}}\,K_0(2mr_0 v) \\ \times \cosh\left[\left(|2F-1| - 1\right){\rm arccosh} v\right] \\
+ \frac{e}{\pi^2} \!\int\limits_{0}^{\pi/2} \!d\xi \!\int\limits_{m
r_0}^{\infty}\!dv\, v  \Biggl\{{\rm sgn}\!\left(F-\frac12\right)C^{(\pm)}_{\frac12+\left|F-\frac12\right|}\left(v, \sqrt{v^2\sin^2\xi+m^2 r_0^2\cos^2\xi}\right)
D_{\frac12+\left|F-\frac12\right|}(v) \Biggr. \\
+ \sum_{l=1}^\infty \Biggl[C^{(\pm)}_{l+F}\left(v, \sqrt{v^2\sin^2\xi+m^2 r_0^2\cos^2\xi}\right) D_{l+F}(v) \Biggr. \\
\Biggl. \Biggl.  - C^{(\pm)}_{l+1-F}\left(v, \sqrt{v^2\sin^2\xi+m^2 r_0^2\cos^2\xi}\right)
D_{l+1-F}(v) \Biggr]\Biggr\}. 
\end{multline}

\begin{figure}[t]
\begin{center}
\includegraphics[width=160mm]{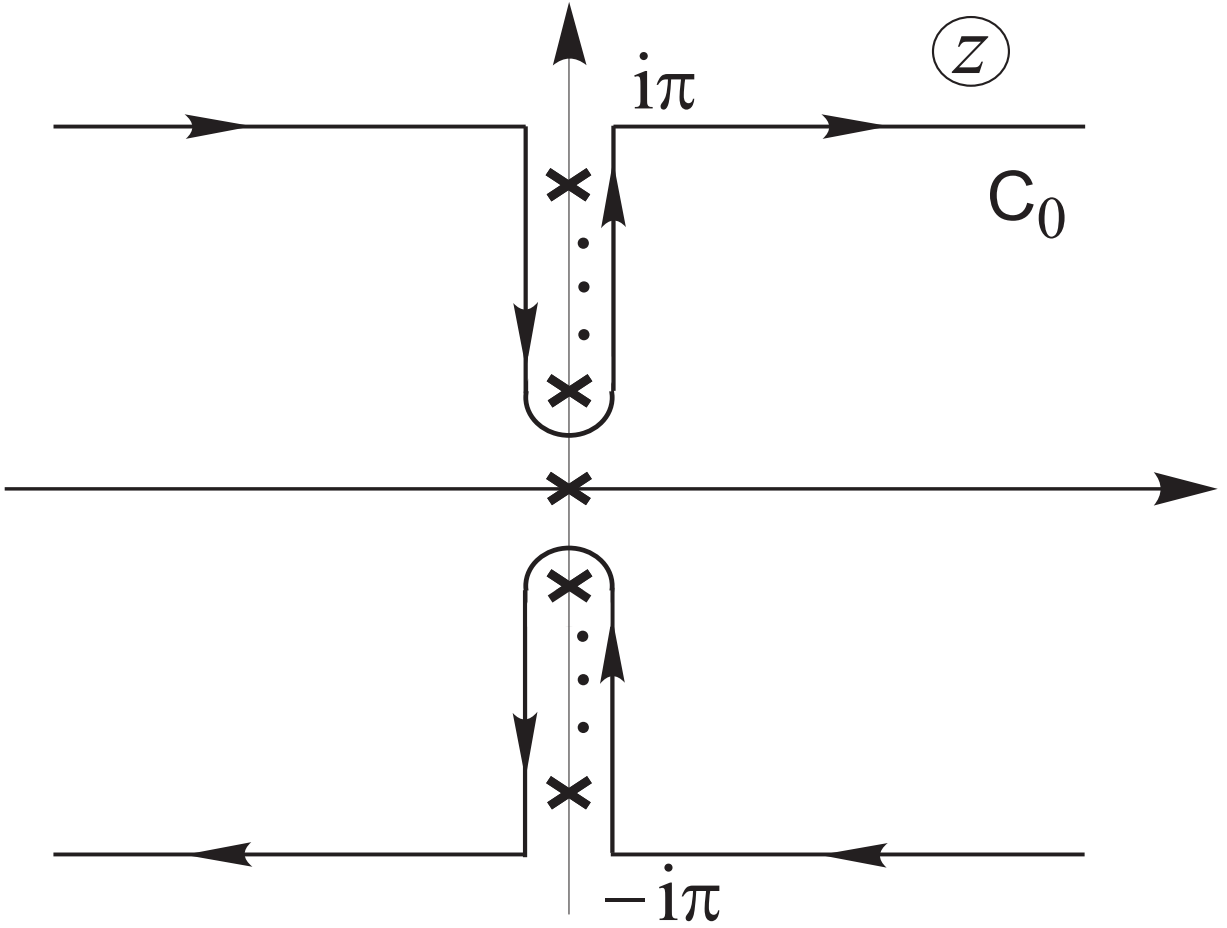}
\end{center} \caption{Contour $C_0$ on complex $z$ plane; the vertical parts of the contour can be infinitesimally close to the ordinate axis. Singularities of the integrand in the last line of \eqref{5.12} are indicated by crosses.}\label{Fig}
\end{figure}

A piece of $\left. \Phi_{\rm I}\right|_{F \neq 1/2, \, |\tau_1|=|\tau_{-1}| = 1}$ with terms containing the $K_0(w)$ function is prevailing at $r_0 \ll m^{-1}$: it corresponds to the contribution of current  $j_\varphi^{(a)(s)}$ defined according to \eqref{4.18}. This piece can be presented as an integral over contour $C_0$ in the complex $z$ plane, see Figure,
\begin{multline}\label{5.12}
\left. \Phi_{\rm I}^{(a)}\right|_{F \neq 1/2, \, |\tau_1|=|\tau_{-1}| = 1} = \frac{e}{2 \pi}\left\{\frac{1}{2\pi} \int\limits_0^\infty
\frac{du}{\cosh^3(u/2)}\,K_0[2mr_0 \cosh(u/2)]\,\Omega_{{\rm sgn}(F - \frac12)}(u)\right. \\
\left.+ \frac{1}{\nu}\sum_{p=1}^{\left[\!\left| {\nu}/2
\right|\!\right]} K_0[2mr_0\sin(p\pi/\nu)]
\,\frac{\sin[(2F-1)p\pi]}{\sin^3(p\pi/\nu)} - \frac{1}{4N} K_0 (2mr_0)
\sin\left[\left(2F -1\right) N \pi \right] \,
\delta_{\nu, \, 2N}\right\} \\ 
= - \frac{e}{8 \pi}\,\frac{{\rm sgn}\left(F-\frac12\right)}{2\pi {\rm
i}}\int\limits_{C_0}
dz\,K_0 \left(2mr_0\sqrt{-\sinh^2(z/2)}\right)\,\frac{\sinh\left[\nu\left(|F-\frac12| - \frac12\right)z\right]}{\sinh^3(z/2)\sinh(\nu
z/2)};
\end{multline}
here the integral over $u$ corresponds to the contribution of the horizontal parts of contour $C_0$, while other terms correspond to the contribution of simple poles at $|{\rm Im} z| \leq  \pi$ on the ordinate axis out of the origin. Using the asymptotics of the $K_0(w)$ function at small values of its argument, one can get
\begin{equation}\label{5.13}
\left. \Phi_{\rm I}\right|_{F \neq 1/2, \, 
|\tau_1|=|\tau_{-1}| = 1, \, m r_0 \ll 1} = - \frac{e}{8 \pi}\,\frac{{\rm sgn}\left(F-\frac12\right)}{2\pi {\rm
i}}\int\limits_{C_0}
dz\,\frac{\sinh\left[\nu\left(|F-\frac12| - \frac12\right)z\right]}{\sinh^3(z/2)\sinh(\nu
z/2)}\left[-\ln (m r_0)\right].
\end{equation}
Now, a singularity at the origin of the complex $z$ plane is just an isolated pole, and contour $C_0$ can be continuously deformed  into a contour encircling the origin. Calculating the residue of a simple pole at $z=0$, we obtain
\begin{multline}\label{5.14}
\left. \Phi_{\rm I}\right|_{F \neq 1/2, \, 
|\tau_1|=|\tau_{-1}| = 1, \, m r_0 \ll 1} \, = \, -\frac{e}{3 \pi} \left[F -\frac12 -
\frac12{\rm sgn}\left(F-\frac12\right)\right] \\ \times \Biggl\{\frac34 -\nu^2
\left[\frac14 - \left|F -\frac12\right| - F(1-F)\right]\Biggr\}\left[-\ln (m r_0)\right]. 
\end{multline}
Thus, the flux which is induced in the vacuum by a cosmic string of small transverse size is discontinuous at $F = 1/2$, and the discontinuity is independent of the string tension: 
\begin{equation}\label{5.15}
\left. \lim_{F\rightarrow (1/2)_{\pm}}\Phi_{\rm I}\right|_{F \neq 1/2, \, |\tau_1|=|\tau_{-1}| = 1, \, m r_0 \ll 1} = \, \pm \, \frac{e}{8 \pi}\left[-\ln (m r_0)\right], 
\end{equation}
this is certainly a consequence of \eqref{4.20} or \eqref{4.21}.

Considering the case of a cosmic string of nonsmall transverse size, we present \eqref{5.7} as, cf. \eqref{4.3},
\begin{equation}\label{5.16}
\left. \Phi_{\rm I}\right|_{F \neq 1/2, \, \tau_1=\tau_{-1} = \pm 1} = \int\limits_0^\infty \frac{dk_3}{\pi} \,\, \left. \Phi^{(2dim)}_{\rm I}(m_3)\right|_{F \neq 1/2, \, 
\tau_1 = \pm 1} ,  \quad  m_3 = \sqrt{m^2+k_3^2}, 
\end{equation}
where
\footnote{Note that $\left. \Phi^{(2dim)}_{\rm I}\right|_{F = 1/2, \, \tau_1 = - 1_{\mp}}$ which is proportional to $m^{-1}$, see (6.32) in \cite{SiG}, yields a divergence of $\left. \Phi_{\rm I}\right|_{F = 1/2, \, \tau_1=\tau_{-1} = - 1}$. In contrast to this, terms that are proportional to $m^{-1}$ in \eqref{5.17} are harmless, yielding terms in the first figure brackets in \eqref{5.7}.}
\begin{multline}\label{5.17}
\left. \Phi^{(2dim)}_{\rm I}(m)\right|_{F \neq 1/2, \, \tau_1 = \pm 1} = \frac{e}{2m}\left\{\frac{1}{2\pi} \int\limits_0^\infty
\frac{du}{\cosh^3(u/2)} \, \exp\left[-2mr_0 \cosh(u/2)\right] \, \Omega_{{\rm sgn}(F - \frac12)}(u)\right. \\
\left.+ \frac{1}{\nu}\sum_{p=1}^{\left[\!\left| {\nu}/2
\right|\!\right]} \exp [-2mr_0\sin(p\pi/\nu)]
\,\frac{\sin[(2F-1)p\pi]}{\sin^3(p\pi/\nu)} - \frac{1}{4N} {\rm e}^{-2mr_0}
\sin\left[\left(2F -1\right) N \pi \right] \,
\delta_{\nu, \, 2N}\right\} \\ 
+ \frac{e}{\pi} r_0 \,\int\limits_{m r_0}^{\infty}
\frac{dv\, v}{\sqrt{v^2 - m^2 r_0^2}} \, \Biggl\{{\rm sgn}\!\left(F-\frac12\right)
C^{(\pm)}_{\lambda\left(0, \, \left|F-\frac12\right|\right)}\left(v, m  r_0\right)
D_{\lambda \left(0, \, \left|F-\frac12\right|\right)}(v) \Biggr. \\
+ \sum_{l=1}^\infty \Biggl[C^{(\pm)}_{\lambda \left(l, \, F-\frac12 \right)}\left(v, m r_0\right) D_{\lambda \left(l, \, F-\frac12 \right)}(v)  
 - C^{(\pm)}_{\lambda \left(l, \, \frac12-F \right)}
\left(v, m r_0\right)
D_{\lambda \left(l, \, \frac12-F \right)}(v) \Biggr]\Biggr\}. 
\end{multline}
The numerical analysis of the latter quantity as a function of $m r_0$ was performed in \cite{SiG}. Whereas $\left. \Phi^{(2dim)}_{\rm I}(m)\right|_{F \neq 1/2, \, \tau_1 = 1}$ decreases in its absolute value with the increase of $m r_0$, $\left. \Phi^{(2dim)}_{\rm I}(m)\right|_{F \neq 1/2, \, \tau_1 = - 1}$ increases at no allowance in its absolute value with the increase of $m r_0$: this is manifest up to values $m r_0 = 10$ for $0.3 < F < 0.7$ and $1 \leq \nu < 2$, although there is a moderate decrease in vicinities of $F=0$ and $F=1$ at $\nu > 2$; certainly, the increase is in general caused by a zero in the denominator in \eqref{5.6} in the case of $\tau_s = - 1$. Thus, we can conclude that $\left. \Phi_{\rm I}\right|_{F \neq 1/2, \, \tau_1=\tau_{-1} = - 1}$ increases at no allowance in its absolute value with the increase of $m r_0$, whereas $\left. \Phi_{\rm I}\right|_{F \neq 1/2, \, \tau_1=\tau_{-1} = 1}$ decreases in its absolute value with the increase of $m r_0$, and 
\begin{equation}\label{5.18}
\left. \lim_{m r_0\rightarrow \infty}\Phi_{\rm I}\right|_{F \neq 1/2, \, \tau_1=\tau_{-1} = 1} = 0.
\end{equation}

\section{Summary}
\setcounter{equation}{0}
\renewcommand{\theequation}{\arabic{section}.\arabic{equation}}

In the present paper, we have shown that a straight cosmic string of nonvanishing transverse size induces a magnetic field in the vacuum of the quantum relativistic charged spinor matter field. Let us recall first that, as was noted in Introduction, the vacuum polarization effects in the cosmic string background depend periodically on the gauge flux of the string, $\Phi$ \eqref{1.3} (this is a consequence of the Aharonov-Bohm effect). The period equals to $2\pi/ \tilde e$ with $\tilde e$ being the coupling constant of the matter field to the gauge field of the string, see \eqref{1.33}, and the effects disappear at $\Phi = 2\pi n/ \tilde e$, where $n \in \mathbb{Z}$. Thus, the dependence is on variable $F$ ranging from $0$ to $1$, and not on $n_c$, see \eqref{2.27}, with $F = 1/2$ corresponding to $\Phi = 2\pi (n + 1/2) / \tilde e$. Of paramount importance is the issue of boundary conditions at the edge of the string core. The most general boundary condition ensures the self-adjointness of the Hamiltonian operator for the matter field and the impenetrability of the string core. Such a condition depends on four parameters with each one depending on a point of the boundary. However, even if the number of parameters is reduced by half, see Section 3, still this generality is excessive and leads to unphysical consequences. Namely, we have obtained analytic expressions for the temporal and spatial components of the induced vacuum current in the case of either $P$ or $CT$ invariant boundary condition, when there are two position-dependent boundary parameters, see Section 4. The total induced vacuum charge is found to be infinite in this case, which is due to a lack of sufficient decrease of the temporal component in the direction of the string axis, and the infinity could hardly be accepted as physically plausible. Therefore the only choice is the $CPT$ invariant boundary condition with one position-dependent boundary parameter, see \eqref{3.34} and \eqref{3.35}: the total induced vacuum charge vanishes in this case (see the ending of Section 4). 

Further restrictions on the boundary condition follow from the analysis of the induced vacuum magnetic field which, by virtue of the Maxwell equation, is related to the spatial current component. The total induced vacuum magnetic flux, $\Phi_I$ \eqref{2.31}, is infinite in the case of the two-parameter position-dependent boundary condition, i.e. when either $P$ or $CT$ invariance holds. It is finite at $F \neq 1/2$, see \eqref{5.7} and \eqref{5.8}, in the case of the $C$ invariant boundary condition,
\begin{equation}\label{6.1}
    (I \mp {\rm i} \gamma^r)\left.\psi\right|_{r=r_0}=0,
\end{equation}
with boundary parameters taking values 
$\tau_1 = \pm 1$ and $\tau_{-1} = \pm 1$. One can conclude that the requirement of the finiteness for $\Phi_I$ at $F \neq 1/2$, being equivalent to the requirement of the charge conjugation invariance, leads to a somewhat ambiguous result (although there are no position-dependent parameters). 

An ambiguity in the value of the total induced vacuum magnetic flux is completely removed by elaborating the case of $F=1/2$. We show that the only way to avoid a divergence of flux $\Phi_I$ at $F=1/2$ is to impose the unambiguous condition,
\begin{equation}\label{6.2}
    (I-{\rm i} \gamma^r)\left.\psi\right|_{r=r_0}=0,
\end{equation}
then $\tau_1 = \tau_{-1} = 1$ and the flux is zero, see \eqref{5.5}. 

Returning to the case of $F \neq 1/2$, we note that $\left. \Phi_{\rm I}\right|_{F \neq 1/2, \,  \tau_1=\tau_{-1}=-1}$ increases at no allowance in its absolute value with the increase of $m r_0$, see the ending of Section 5. Such a behavior is hardly to be regarded as physically plausible, if the transverse size of the string is somehow identified with the correlation length, see \eqref{1.5}. Really, it looks rather unlikely that a topological defect (cosmic string) influences the surrounding quantum matter with the matter particle mass, $m$, exceeding the energy scale of spontaneous symmetry breaking, $m_{\rm H}$; the more unlikely is the unrestricted growth of this influence with the increase of quotient $m/m_{\rm H}$. The influence of a topological defect on the surrounding quantum matter at $m_{\rm H} \gg m$ (which is the right inequality in \eqref{1.4}) looks much more physically plausible. Namely this situation is realized for $\left. \Phi_{\rm I}\right|_{F \neq 1/2, \,  \tau_1=\tau_{-1}=1}$ (note, in particular, relation \eqref{5.18}) and in the case of the quantum charged scalar matter field obeying the Dirichlet boundary condition at the edge of the string core, see \cite{Gor2,Gor3,Gor4,Gor1}.  

Thus, we conclude that the requirement of the physically plausible behavior for the induced vacuum magnetic flux results in the unambiguous determination of the boundary condition for the quantum charged spinor matter field at the edge of the string core. The condition coincides with the renowned quark bag boundary condition \cite{Bog,Cho1,Joh}. Apparently, the authors of \cite{Bog,Cho1,Joh} were lucky actually to guess the proper boundary condition, and this has been demonstrated in the present study. We list here the unambiguously determined expressions for the induced vacuum magnetic field strength in the cosmic string background, which are valid in the case of $1 \leq \nu < \infty$ and in the case of $\frac12 \leq \nu < 1$ for $\frac12 \left(\frac1\nu - 1\right) < F < \frac12 \left(3 - \frac1\nu\right)$:
\begin{multline}\label{6.3}
\left.B_I(r)\right|_{F \neq 1/2, \, \tau_{1}=\tau_{-1}=1}  
=\frac{e \nu m}{2 \pi^2 r}\left\{ \frac{1}{2\pi}\int\limits_0^\infty
\frac{du}{\cosh^2 (u/2)}K_1\left[2mr \cosh(u/2)\right]\, \Omega_{{\rm sgn}\left(F - \frac12\right)}(u)\right. \\
\left.+\frac{1}{\nu}\sum\limits_{p=1}^{\left[\!\left| {\nu}/2 \right|\!\right]}K_1\left[2mr\sin(p\pi/\nu)\right]\frac{\sin\left[(2F-1)p\pi\right]}{\sin^2(p\pi/\nu)}-\frac{1}{4N}K_1(2mr)\sin\left[(2F-1)N\pi\right]\delta_{\nu,2N}\right\}\\
-\frac{e \nu r}{\pi^3}\!\int\limits_0^{\pi/2}\!\!d{\xi}\!\int\limits_m^\infty\!\!dq\, q^2 
\Biggl\{{\rm sgn}\!\left(F-\frac12\right)C^{(+)}_{\lambda \left(0, \, \left|F-\frac12\right| \right)}\left(qr_0, \sqrt{q^2\sin^2\xi+m^2\cos^2\xi}\,\,r_0\right)  W_{\lambda \left(0, \, \left|F-\frac12\right|\right)}(qr) \Biggr. \\
+ \sum_{l=1}^{\infty} \Biggl[ C^{(+)}_{\lambda \left(l, \, F-\frac12 \right)}\left(qr_0, \sqrt{q^2\sin^2\xi+m^2\cos^2\xi}\,\,r_0\right)
W_{\lambda \left(l, \, F-\frac12 \right)} (qr)
\Biggr. \\
\Biggl.\Biggl. - C^{(+)}_{\lambda \left(l, \, \frac12-F \right)}\left(qr_0, \sqrt{q^2\sin^2\xi+m^2\cos^2\xi}\,\,r_0\right) W_{\lambda \left(l, \, \frac12-F \right)}(qr)\Biggr]\Biggr\},
\end{multline}
where $\Omega_{\pm}(u)$ is given by \eqref{5.10}, $C^{(+)}_\rho(v, w)$ is given by  \eqref{5.6}, $W_{\rho}(v)$ is given by \eqref{4.112}, $\lambda(l, y)$ is given by \eqref{4.44}, and  
\begin{equation}\label{6.4}
\left.B_I(r)\right|_{F = 1/2, \, \tau_{1}=\tau_{-1}=1} = 0. 
\end{equation}
Note that the strength changes sign under substitution $F \rightarrow 1-F$, vanishing at $F = 0, 1/2, 1$, and decreases at large distances from the string, $r \rightarrow \infty$, as
$$\left\{
\begin{array}{l}
r^{-2}\,\, {\rm exp}(-2mr), \quad \frac12 \leq \nu < 2  \\
\vphantom{\int\limits_0^0}
m^{1/2} r^{-3/2}\,\, {\rm exp}[-2mr\sin(\pi/\nu)], \quad \nu\geq 2
\end{array}
\right\},
$$
while it behaves as $(r-r_0)^{-1 + \varepsilon}$ with $\varepsilon > 0$  at $r \rightarrow r_0$, see \eqref{5.2}. Thereby, a cosmic string is enclosed in a sheath in the form of a tube of the magnetic flux lines along the string; the transverse size of the magnetic tube is of the order of the Compton wavelength of the matter particle, $m^{-1}$, and the latter exceeds the transverse size of the string, which is of the order of the correlation length (see \eqref{1.5} and the right inequality in \eqref{1.4}). The total induced vacuum magnetic flux is given by expression \eqref{5.7} with the ``$+$'' sign.  

Although the limit of the vanishing transverse size of a cosmic string can be regarded to be purely formal, it, as unambiguously defined in Appendix C, yields the prevailing contribution in the realistic case of the Compton wavelength of the matter particle exceeding considerably the correlation length. In the $r_0 = 0$ case, the induced vacuum magnetic field strength is given by \eqref{c5}, vanishing at $F = 1/2$; note a discontinuity of the strength at $F = 1/2$, see \eqref{c12}. In the case of a cosmic string of small transverse size, $r_0 \ll m^{-1}$, the total induced vacuum magnetic flux is given by \eqref{5.14}; it is discontinuous at $F = 1/2$, and the discontinuity is independent of the string tension, see \eqref{5.15}. Note that the discontinuity of the results at $F = 1/2$ in the $r_0 \ll m^{-1}$ case is due to the appearance of the peculiar mode in the solution to the Dirac equation [see \eqref{a5} in Appendix A and \eqref{c3} in Appendix C]. It would be interesting to consider other induced vacuum effects, for instance, the induced vacuum energy-momentum tensor in the cosmic string background.

\section*{Acknowledgments}
The work was supported by the National Academy of
Sciences of Ukraine (Project No.01172U000237) and by the Program of
Fundamental Research of the Department of Physics and Astronomy of
the National Academy of Sciences of Ukraine (Project No.0117U000240).


\setcounter{equation}{0}
\renewcommand{\theequation}{A.\arabic{equation}}
\section*{Appendix A: Solution to the Dirac equation}

The solution to the system of equations, \eqref{2.25}, is given in
terms of cylindrical functions. Let us define
\begin{multline}\label{a1}
\left(
\begin{array}{c}
f_n^{(\wedge)(s)} \\ g_n^{(\wedge)(s)}
\end{array}
\right)= \frac12 \sqrt{\frac{\nu}{\pi} } \\
\times \left(
\begin{array}{c}
\sqrt{1+m_3/E} \left[\sin(\mu^{(\wedge)(s)}_{\nu l +1 -G_s}) J_{\nu l-G_s}(kr) + \cos(\mu^{(\wedge)(s)}_{\nu l +1 -G_s}) Y_{\nu l-G_s}(kr)\right] \\
{\rm sgn}(E) \sqrt{1- m_3/E} \left[\sin(\mu^{(\wedge)(s)}_{\nu l +1 -G_s})
J_{\nu l+1-G_s}(kr) + \cos(\mu^{(\wedge)(s)}_{\nu l +1 -G_s}) Y_{\nu
l+1-G_s}(kr)\right]
\end{array}
\right),
\end{multline}
where $l=s (n-n_{\rm c})$, and
\begin{multline}\label{a2}
\left(
\begin{array}{c}
f_n^{(\vee)(s)} \\ g_n^{(\vee)(s)}
\end{array}
\right)=  \frac12 \sqrt{\frac{\nu}{\pi} } \\
\times \left(
\begin{array}{c}
\sqrt{1+m_3/E} \left[\sin(\mu^{(\vee)(s)}_{\nu l' +G_s}) J_{\nu l'+G_s}(kr) + \cos(\mu^{(\vee)(s)}_{\nu l' +G_s}) Y_{\nu l'+G_s}(kr)\right] \\
-{\rm sgn}(E) \sqrt{1- m_3/E} \left[\sin(\mu^{(\vee)(s)}_{\nu l' +G_s})
J_{\nu l'-1+G_s}(kr) + \cos(\mu^{(\vee)(s)}_{\nu l' +G_s}) Y_{\nu
l'-1+G_s}(kr)\right]
\end{array}
\right),
\end{multline}
where $l'=-s (n-n_{\rm c})$; here $J_\rho(u)$ and  $Y_\rho(u)$ are the Bessel
and Neumann functions of order $\rho$, 
$k = \sqrt{E^2 - m^2_3}$.

In the case of $\nu > 1$ and $-\frac12 < s\left(F -\frac12\right) < -\frac1{2\nu}$
$\quad$ $\left( \frac12 ( 1-\nu ) < G_s <0 \right)$, the complete set of solutions to \eqref{2.25} is given by
\begin{equation}\label{a3}
\left.
\left( \begin{array}{c} f_n^{(s)} \\ g_n^{(s)}
\end{array}
\right)\right|_{s (n-n_{\rm c}) \geq 0} = \left(
\begin{array}{c}
f_n^{(\wedge)(s)} \\ g_n^{(\wedge)(s)}
\end{array}
\right), \quad
\left. \left( \begin{array}{c} f_n^{(s)} \\ g_n^{(s)}
\end{array}
\right)\right|_{s (n-n_{\rm c}) \leq -1} = \left(
\begin{array}{c}
f_n^{(\vee)(s)} \\ g_n^{(\vee)(s)}
\end{array}
\right).
\end{equation}
In the case of $\nu >1 $ and $\frac1{2\nu} < s\left(F -\frac12\right) < \frac12$
$\quad$ $\left( 1<G_s < \frac12(1+\nu)\right)$, the complete set of solutions to \eqref{2.25} is given by
\begin{equation}\label{a4}
\left. \left( \begin{array}{c} f_n^{(s)} \\ g_n^{(s)}
\end{array}
\right)\right|_{s (n-n_{\rm c}) \geq  1} = \left(
\begin{array}{c}
f_n^{(\wedge)(s)} \\ g_n^{(\wedge)(s)}
\end{array}
\right), \quad \left. \left( \begin{array}{c} f_n^{(s)} \\ g_n^{(s)}
\end{array}
\right)\right|_{s (n-n_{\rm c}) \leq 0} = \left(
\begin{array}{c}
f_n^{(\vee)(s)} \\ g_n^{(\vee)(s)}
\end{array}
\right).
\end{equation}

In the case of $\nu \geq 1$ and $|F-\frac12| < \frac{1}{2\nu}$ $\quad$ $(0<G_s<1)$,
there is a peculiar mode corresponding to $n=n_{\rm c}$; the mode can be chosen in the following form:
\begin{multline}\label{a5}
\left(
\begin{array}{c}
f_{n_{\rm c}}^{(s)} \\ g_{n_{\rm c}}^{(s)}
\end{array}
\right)= \frac12 \sqrt{\frac{\nu}{\pi} }
\frac1{\sqrt{1+\sin(2\mu_{1-G_s}^{(c)(s)})\cos(G_s\pi) } } \\ \times \left(
\begin{array}{c}
\sqrt{1+m_3/E} \left[\sin(\mu_{1 -G_s}^{(c)(s)}) J_{-G_s}(kr) + \cos(\mu_{1 -G_s}^{(c)(s)}) J_{G_s}(kr)\right] \\
{\rm sgn}(E) \sqrt{1- m_3/E} \left[\sin(\mu_{1 -G_s}^{(c)(s)}) J_{1-G_s}(kr) -
\cos(\mu_{1 -G_s}^{(c)(s)}) J_{-1+G_s}(kr)\right]
\end{array}
\right).
\end{multline}
Modes
\begin{equation}\label{a6}
\left. \left( \begin{array}{c} f_n^{(s)} \\ g_n^{(s)}
\end{array}
\right)\right|_{s (n-n_{\rm c}) \geq  1} = \left(
\begin{array}{c}
f_n^{(\wedge)(s)} \\ g_n^{(\wedge)(s)}
\end{array}
\right), \quad \left. \left( \begin{array}{c} f_n^{(s)} \\ g_n^{(s)}
\end{array}
\right)\right|_{s (n-n_{\rm c}) \leq -1} = \left(
\begin{array}{c}
f_n^{(\vee)(s)} \\ g_n^{(\vee)(s)}
\end{array}
\right)
\end{equation}
together with mode \eqref{a5} comprise the set of all solutions with $|E|>m_3$ in this case.

In the case of $\frac12 \leq \nu< 1$ and $|F-\frac12| < 1 - \frac{1}{2\nu}$ $\quad$ $(1-\nu < G_s <\nu)$, the set of all solutions with $|E|>m_3$ is also given by \eqref{a5} and \eqref{a6}. In the case of $\frac12 \leq \nu < 1$ and either $-\frac12 < s\left(F -\frac12\right) < - 1 + \frac1{2\nu}$ 
$\quad$ $\left( \frac12(1-\nu) < G_s < 1-\nu \right)$ or 
$ 1 - \frac{1}{2\nu} < s\left(F -\frac12\right) < \frac12$ 
$\quad$ $\left( \nu < G_s < \frac12(1+\nu) \right)$, there are two peculiar modes. In the case of $0< \nu < \frac12$, there are two and more peculiar modes.

Certainly, the limit of $r\rightarrow 0$ is of no sense for cosmic strings
of nonzero transverse size. However, it is instructive here to discuss briefly the case of an infinitely  thin cosmic string (see \cite{SiG} for details). Most of the modes in the $r_0 = 0$ case are obtained by
putting $\mu^{( \wedge )(s)}_\rho = \mu^{( \vee )(s)}_\rho =\pi/2$ in \eqref{a3}, \eqref{a4}, and \eqref{a6}; these modes are regular at $r\rightarrow 0$.
However, peculiar mode \eqref{a5} cannot be
made regular at $r \rightarrow 0$; it is irregular but square
integrable. The latter circumstance requires a quest for a
self-adjoint extension, and the Weyl-von Neumann theory of
deficiency indices (see \cite{Ree,Alb}) has to be employed. In the case of
$\nu \geq 1$ and $|F-\frac12| < \frac{1}{2\nu}$, as well as in the case of $\frac12 \leq \nu < 1$ and $|F-\frac12| < 1 - \frac{1}{2\nu}$, when there is one irregular mode,
the deficiency index is (1,1), and the one-parameter family of self-adjoint extensions is introduced. The induced vacuum current and other vacuum polarization effects in 
two-dimensional surface $z={\rm const}$ were comprehensively and exhaustively studied for $\nu=1$ in \cite{Sit6,SiR,Si7,Sit9,Si9} and for carbonlike nanocones in \cite{SiV7,SiV1,SiV2,Si18}. 
In the case of $\frac12 \leq \nu<1$ and either $-\frac12 < s\left(F -\frac12\right) < - 1 + \frac1{2\nu}$  or 
$ 1 - \frac{1}{2\nu} < s\left(F -\frac12\right) < \frac12$, and other cases, when there are two irregular square integrable modes, the deficiency index is (2,2), and there are four self-adjoint extension parameters. These cases remain unstudied yet.

Imposing the $P$ invariant boundary condition with matrix $K$ \eqref{3.10} or the $CT$ invariant boundary condition with matrix $K$  \eqref{3.24} on the solution to the Dirac equation,
$\psi_E(\textbf{x})$ \eqref{2.17}, we obtain the condition for the modes, see \eqref{3.19} or \eqref{3.29}, which
allows us to determine the mode coefficients:
\begin{align}
& \tan(\mu^{(\wedge)(s)}_\rho) = \frac{k Y_{\rho-1}(kr_0)- \tau_s (m_3-E) Y_{\rho}(kr_0)}
{- k J_{\rho-1}(kr_0)+ \tau_s (m_3-E) J_{\rho}(kr_0)},  \label{a7}\\
& \tan(\mu^{(\vee)(s)}_\rho  ) =
\frac{k Y_{\rho-1}(kr_0)- \tau_s^{-1} (m_3+E) Y_{\rho}(kr_0)}
{- k J_{\rho-1}(kr_0)+ \tau_s^{-1} (m_3+E) J_{\rho}(kr_0)},  \label{a8}\\
& \tan(\mu_{1-G_s}^{(c)(s)}) = \frac{k J_{G_s}(kr_0)+ \tau_s (m_3 - E) J_{-1+G_s}(kr_0)}
{- k J_{-G_s}(kr_0) + \tau_s (m_3-E) J_{1-G_s}(kr_0)};
\label{a9}
\end{align}
note that relation $\tau_{-s} = \tau_s^{-1}$ holds in the case of the $CPT$ invariant boundary condition, and, consequently, we obtain
\begin{equation}\label{a10}
\tan(\mu^{(\vee)( - s)}_\rho  ) = \left.\tan(\mu^{(\wedge)(s)}_\rho)\right|_{E \rightarrow - E} 
\end{equation}
and
\begin{equation}\label{a11}
 \tan(\mu_{G_s}^{(c)(-s)}) = - \left.\cot(\mu_{1-G_s}^{(c)(s)})\right|_{E \rightarrow - E}
\end{equation}
in this case.

Because of conditions \eqref{3.19} and \eqref{3.29} in general case, in addition to the continuous spectrum, a bound state with energy $E_{BS}^{(s)}$ in the gap between the continuums, $-m_3 < E_{BS}^{(s)}< m_3$, emerges in section $z={\rm const}$ at $\tau_s < 0$ for $n=n_{\rm c}$
$\quad$ ($\nu \geq 1$ and $|F-\frac12| < \frac{1}{2\nu}$, or $\frac12
\leq \nu <1$ and $|F-\frac12| < 1 - \frac{1}{2\nu}$). Its mode is 
\begin{multline}\label{a12}
\left( \begin{array}{c} f_{n_{\rm c}}^{(BS)(s)} \vphantom{\int\limits_0^0}\\
g_{n_{\rm c}}^{(BS)(s)}
\end{array}
\right) = \sqrt{\frac{\nu\kappa_s m_3}{2 \pi r_0}}\Biggl\{ m_3 K_{G_s}(\kappa_s r_0)
K_{1-G_s}(\kappa_s r_0)  \Biggr. \\
\Biggl.+E_{BS}^{(s)}\Biggl[\kappa_s r_0 K^2_{1-G_s}(\kappa_s r_0)-\kappa_s r_0 K^2_{G_s}(\kappa_s r_0)+ (2G_s-1)K_{G_s}(\kappa_s r_0) K_{1-G_s}(\kappa_s r_0) \Biggr]
\Biggr\}^{-1/2} \\ \times \left(
\begin{array}{c} \sqrt{1+E_{BS}^{(s)}/m_3} \,K_{G_s}(\kappa_s r) \\  \sqrt{1-E_{BS}^{(s)}/m_3} \,K_{1-G_s}(\kappa_s r)
\end{array}
\right),
\end{multline}
where $\kappa_s = \sqrt{m^2_3-(E_{BS}^{(s)})^2}$, and the value of its energy is determined from relation
\begin{equation}\label{a13}
\sqrt{ \frac{1+E_{BS}^{(s)}/m_3}{1-E_{BS}^{(s)}/m_3 }} = - \tau_s \, \frac{K_{1-G_s}(\kappa_s r_0)}{K_{G_s}(\kappa_s r_0)}.
\end{equation}

Comparing the case of a cosmic string of nonzero transverse size with that of an infinitely thin one, we emphasise that in the first case all partial Hamiltonians are extended with the same boundary parameters, whereas in the second case several partial Hamiltonians are extended, and the number of self-adjoint extension parameters can be zero (no need for extension, the operator is essentially self-adjoint), one, four, etc. The values of the self-adjoint extension parameters in the second case can be fixed from the first case by limiting procedure $r_0\rightarrow 0$ which transforms peculiar modes into irregular ones. Namely in this way, the condition of minimal irregularity
\cite{Sit6,Si7} is obtained in the case of the deficiency
index  equal to (1,1), i.e. when only one peculiar mode exists.

\setcounter{equation}{0}
\renewcommand{\theequation}{B.\arabic{equation}}
\section*{Appendix B: Extracting the dependence on the transverse size of a cosmic string} 

In the situation when peculiar modes are absent, i.e. $\nu \geq 1$ and $\frac{1}{2\nu} < |F-\frac12| < \frac{1}{2}$, all the $r_0$ dependence is contained in integrals over $q$ in \eqref{4.12}-\eqref{4.15}. We are considering below the situation when there is one peculiar mode for each value of $s$, i.e. $\nu \geq 1$ and $|F-\frac12| < \frac{1}{2\nu}$, or $\frac12 \leq \nu <1$ and $|F-\frac12| < 1 - \frac{1}{2\nu}$.

In the case of $F\neq 1/2$, we present 
$j_\varphi^{(s)}(r)$ as
\begin{multline}\label{b1}
\left.j_{\varphi}^{(s)}(r)\right|_{F\neq 1/2, \,(\tau_s)^{-s{\rm sgn}(F-\frac{1}{2})}\neq 0}\\
=\frac{m^2}{2 \pi^2}\left\{\frac{1}{2\pi}\int\limits^{\infty}_{0}\frac{du}{\cosh(u/2)}K_2\left[2mr\cosh(u/2)\right]\Omega_{{\rm sgn}(F-\frac{1}{2})}(u)\right.\\ 
\left. +\frac{1}{\nu}\sum\limits_{p=1}^{\left[\!\left| {\nu}/2 \right|\!\right]} K_2\left[2mr\sin(p\pi/\nu)\right]\frac{\sin\left[(2F-1)p\pi\right]}{\sin(p\pi/\nu)}-\frac{1}{4N}K_2(2mr)\sin\left[(2F-1)N\pi\right]\delta_{\nu,2N}\right\}\\ 
+\frac{sr}{\pi^3}\int\limits_{0}^{\pi/2}d\xi\int\limits_{m}^{\infty}dqq^2\left\{\Biggl[\Theta\!\left(sF-\frac{s}{2}\right)
C^{(\vee)(s)}_{\lambda\left[0, \, s\left(F-\frac12\right)\right]}\left(qr_0,\sqrt{q^2\sin^2\xi+m^2\cos^2\xi} \, r_0\right)\Biggr.\right. \\ 
\Biggl. - \Theta\!\left(\frac{s}{2}-sF\right) 
C^{(\wedge)(s)}_{\lambda\left[0, \, s\left(\frac12-F\right)\right]}\left(qr_0,\sqrt{q^2\sin^2\xi+m^2\cos^2\xi} \, r_0\right)\Biggr] K_{\lambda \left(0, \, F-\frac12 \right)}(qr)K_{\lambda \left(0, \, \frac12-F \right)}(qr)\\
+\sum\limits_{l=1}^{\infty}\left[C_{\lambda\left[l, \, s\left(F-\frac12\right)\right]}^{(\vee)(s)}\left(qr_0,\sqrt{q^2\sin^2\xi+m^2\cos^2\xi} \, r_0\right)K_{\lambda\left[l, \, s\left(F-\frac12\right)\right]}(qr)K_{\lambda\left[l, \, s\left(F-\frac12\right)\right]-1}(qr)\right.\\ 
\Biggl.\left.-C_{\lambda\left[l, \, s\left(\frac12-F\right)\right]}^{(\wedge)(s)}\left(qr_0,\sqrt{q^2\sin^2\xi+m^2\cos^2\xi} \,r_0\right)K_{\lambda\left[l, \, s\left(\frac12-F\right)\right]}(qr)K_{\lambda\left[l, \, s\left(\frac12-F\right)\right]-1}(qr)\right]\Biggr\}
\end{multline}
and
\begin{multline}\label{b2}
\left.j_{\varphi}^{(s)}(r)\right|_{F\neq 1/2, \, \ln|\tau_s|^s=\pm\infty}
=\frac{m^2}{2 \pi^2}\left\{\frac{1}{2\pi}\int\limits_{0}^{\infty} \frac{du}{\cosh(u/2)} K_2\left[2mr\cosh(u/2)\right] {\Omega}_{\mp}(u) \right. \\
\left. +\frac{1}{\nu}\sum\limits_{p=1}^{\left[\!\left| {\nu}/2 \right|\!\right]} K_2\left[2mr\sin(p\pi/\nu)\right]\frac{\sin\left[(2F-1)p\pi\right]}{\sin(p\pi/\nu)} - \frac{1}{4N} K_2(2mr)\sin[(2F-1)N\pi]\delta_{\nu,2N}\right\} \\ 
\mp\frac{r}{2 \pi^2}\int\limits_{m}^{\infty}dqq^2\left\{\frac{I_{\lambda\left[0, \, \mp\left(F-\frac12\right)\right]}(qr_0)}{K_{\lambda\left[0, \, \mp\left(F-\frac12\right)\right]}(qr_0)}K_{\lambda \left(0, \, F-\frac12 \right)}(qr) K_{\lambda \left(0, \, \frac12-F \right)}(qr) \right.\\ 
 + \sum\limits_{l=1}^{\infty}\left[\frac{I_{\lambda \left(l, \, F-\frac12 \right)-\frac 12\mp\frac 12}(qr_0)}{K_{\lambda \left(l, \, F-\frac12 \right)-\frac 12\mp\frac 12}(qr_0)}K_{\lambda \left(l, \, F-\frac12 \right)}(qr)K_{\lambda \left(l, \, F-\frac12 \right)-1}(qr) \right. \\ 
\left.\Biggl. + \frac{I_{\lambda \left(l, \, \frac12-F \right)-\frac 12 \pm\frac 12}(qr_0)}{K_{\lambda \left(l, \, \frac12-F \right)-\frac 12\pm\frac 12}(qr_0)}K_{\lambda \left(l, \, \frac12-F \right)}(qr)K_{\lambda \left(l, \, \frac12-F \right)-1}(qr)\right]\Biggr\},
\end{multline}
where $\Theta (t) = \frac12 + \frac12{\rm sgn}(t)$ is the step function, $\Omega_{\pm}(u)$ is given by \eqref{5.10} in Section 5, and the use is made of relations
\begin{equation}\label{b111}
C_{G}^{(\vee)(s)}(v,w) + C_{1 - G}^{(\wedge)(s)}(v,w) = - \frac{2}{\pi}\sin(G\pi) 
\end{equation}
and
\begin{equation}\label{b112}
\!\int\limits_{m}^\infty\!\! dq \, q^2 \, K_{G}(qr) K_{1-G}(qr)\!=\!\frac{m^2}{2r}\!\!\int\limits_0^\infty
\!\!\frac{du}{\cosh(u/2)} \,
\cosh\left[\left(G\!-\!\frac12\right)u\right] K_2\left[2mr\cosh(u/2)\right];
\end{equation}
all the $r_0$ dependence is contained in integrals over $q$ in \eqref{b1} and \eqref{b2}.

In the case of $F=1/2$, we obtain the following expression for the current in two-dimensional surface $z={\rm const}$, see \eqref{4.4}:
\begin{multline}\label{b3}
\left.j_{\varphi}^{(2 {\rm dim})(s)}(r)\right|_{F=1/2}=\frac{s}{2 \pi^2}\left(\tau_{s}^{-1}-\tau_s\right)\int\limits_{m}^{\infty}\frac{dq\,q^2}{\sqrt{q^2-m^2}}\Biggl[\frac{{\rm e}^{2q(r_0-r)}}{q(\tau_{s}^{-1}+\tau_s)+2m}\Biggr. \\ \left.+2r\sum\limits_{l=1}^{\infty}C_{\nu l+\frac 12}^{(s)}\left(qr_o,\,mr_0\right)K_{\nu l+\frac 12}(qr) K_{\nu l-\frac 12}(qr)\right],
\end{multline}
where
\begin{multline}\label{b4}
C_{\nu l+\frac 12}^{(s)}(v,w)\equiv\frac{1}{\tau_{s}^{-1}-\tau_{s}}\left[C_{\nu l+\frac 12}^{(\vee)(s)}(v,w)-C_{\nu l+\frac 12}^{(\wedge)(s)}(v,w)\right] \\ 
=\left\{w\left[K_{\nu l+\frac 12}^{2}(v)+K_{\nu l-\frac 12}^{2}(v)\right]+(\tau_{s}^{-1}+\tau_{s})vK_{\nu l+\frac 12}(v)K_{\nu l-\frac 12}(v)\right\}\\ 
\times\left\{v\left[K_{\nu l+\frac 12}^{2}(v)+K_{\nu l-\frac 12}^{2}(v)\right]\left[v\left(K_{\nu l+\frac 12}^{2}(v)+K_{\nu l-\frac 12}^{2}(v)\right)+2(\tau_{s}^{-1}+\tau_{s})wK_{\nu l+\frac 12}(v)K_{\nu l-\frac 12}(v)\right]\right.\\
\left.+\left[(\tau_{s}^{-1}-\tau_{s})^2v^2+4w^2\right]K_{\nu l+\frac 12}^{2}(v)K_{\nu l-\frac 12}^{2}(v)\right\}^{-1},
\end{multline}
and the use is made of explicit forms,
\begin{equation}\label{b5}
C_{1/2}^{(\vee)(s)}(v,w) - C_{1/2}^{(\wedge)(s)}(v,w) =\frac{2\left(\tau_{s}^{-1}-\tau_s\right){\rm e}^{2v}/{\pi}}{\tau_{s}^{-1}+\tau_{s}+2w/v},\ \ K_{1/2}(v) ={\rm e}^{-v}\sqrt{\frac{\pi}{2v}}.
\end{equation}
Note that, whereas \eqref{b3} vanishes at $\tau_s=1$, it is nonvanishing and discontinuous at $\tau_s=-1$,
\begin{equation}\label{b6}
\left.\lim\limits_{(-\tau_s)^s\rightarrow 1_\mp}j_{\varphi}^{(2{\rm dim})(s)}(r)\right|_{F=1/2}=\pm\frac{m}{2\pi}{\rm e}^{2m(r_0-r)}.
\end{equation}
Inserting \eqref{b3} into \eqref{4.3} and performing integration over $\xi$ in the term corresponding to the first one in square brackets in \eqref{b3}, we obtain
\begin{multline}\label{b7}
\left.j_{\varphi}^{(s)}(r)\right|_{F=1/2}=\frac{s m^2}{\pi^2} \, \frac{{\rm sgn}\left(1-\tau_{s}^2\right)}{(\tau_{s}^{-1}+\tau_{s})^2}\left[\frac{|\tau_{s}^{-1}+\tau_{s}|}{4m(r-r_0)}K_1\left(\frac{4m(r-r_0)}{|\tau_{s}^{-1}+\tau_{s}|}\right) \right. \\
\left. + K_0\left(\frac{4m(r-r_0)}{|\tau_{s}^{-1}+\tau_{s}|}\right)\right] + \frac{s}{4 \pi^3}(\tau_{s}^{-1}-\tau_{s})\int\limits_{m}^{\infty}dq q\Biggl\{{\rm e}^{2q(r_0-r)}\Biggl[K\left(1-m^2/q^2\right)\Biggr.\Biggr.\\ 
\left.-\left(\frac{1+\tau_s^2}{1-\tau_s^2}\right)^2 \Pi\left(-\frac{4(1-m^2/q^2)}{(\tau_{s}^{-1}-\tau_s)^2}; 1 - m^2/q^2\right)\right]\\
\left.+4qr \int\limits_{0}^{\pi/2}d\xi\sum\limits_{l=1}^{\infty}C_{\nu l+\frac 12}^{(s)}\left(qr_0,\sqrt{q^2\cos^2\xi+m^2\sin^2\xi} \, r_0\right)K_{\nu l+\frac{1}{2}}(qr) K_{\nu l-\frac 12}(qr)\right\},
\end{multline}
where
$$
K(m)=\int\limits_{0}^{1} \frac{dt}{\sqrt{(1-t^2)(1-mt^2)}}
$$
is the complete elliptic integral of the first kind, and 
$$
\Pi(n; m)=\int\limits_{0}^{1} \frac{dt}{(1-nt^2)\sqrt{(1-t^2)(1-mt^2)}}
$$
is the complete elliptic integral of the third kind, see \cite{Abra}. In view of relation
\begin{equation}\label{b8}
\lim\limits_{|\tau_s| \rightarrow 1} \frac{4}{|\tau_{s}^{-1}-\tau_s|}\Pi\left(-\frac{4(1-m^2/q^2)}{(\tau_{s}^{-1}-\tau_s)^2}; 1-m^2/q^2\right) = \frac{\pi q}{\sqrt{q^2-m^2}},
\end{equation}
we get
\begin{equation}\label{b9}
\left.j_{\varphi}^{(s)}(r)\right|_{F=1/2, \, \tau_s=1}=0
\end{equation}
and
\begin{multline}\label{b10}
\left.\lim\limits_{(-\tau_s)^s\rightarrow 1_{\mp}}j_{\varphi}^{(s)}(r)\right|_{F=1/2}=\pm \frac{1}{2 \pi^2}\int\limits_{m}^{\infty}\frac{dqq^2}{\sqrt{q^2-m^2}}{\rm e}^{2q(r_0-r)}\\ 
=\pm\frac{m^2}{2 \pi^2}\left\{\frac{1}{2m(r-r_0)}K_1\left[2m(r-r_0)\right]+K_0\left[2m(r-r_0)\right]\right\}.
\end{multline}

Inserting \eqref{b1}, \eqref{b2}, and \eqref{b7} into \eqref{2.30}, we obtain expressions
\begin{multline}\label{b11}
\left.B_I^{(s)}(r)\right|_{F\neq 1/2, \, (\tau_s)^{-s {\rm sgn}(F-\frac 12)}\neq 0} 
\\ 
=\frac{e\nu m}{(2\pi)^2r}\left\{\frac{1}{2\pi}\int\limits_{0}^{\infty}\frac{du}{\cosh^2(u/2)}K_1\left[2mr\cosh(u/2)\right]\Omega_{{\rm sgn}(F-\frac{1}{2})}(u) \right.\\ 
\left.+\frac{1}{\nu}\sum\limits_{p=1}^{\left[\!\left| {\nu}/2 \right|\!\right]}K_1\left[2mr\sin(p\pi/\nu)\right]\frac{\sin\left[(2F-1)p\pi\right]}{\sin^2(p\pi/\nu)}-\frac{1}{4N}K_1(2mr)\sin\left[(2F-1)N\pi\right]\delta_{\nu,2N}\right\}\\ 
-\frac{s e \nu r}{2\pi^3} \int\limits_{0}^{\pi/2}d\xi \int\limits_{m}^{\infty}dq q^2\left\{\Biggl[\Theta\!\left(sF-\frac{s}{2}\right)
C^{(\vee)(s)}_{\lambda\left[0, \, s\left(F-\frac12\right)\right]}\left(qr_0,\sqrt{q^2\sin^2\xi+m^2\cos^2\xi} \,   r_0\right)\Biggr.\right. \\ 
\Biggl. - \Theta\!\left(\frac{s}{2}-sF\right) 
C^{(\wedge)(s)}_{\lambda\left[0, \, s\left(\frac12-F\right)\right]}\left(qr_0,\sqrt{q^2\sin^2\xi+m^2\cos^2\xi} \, r_0\right)\Biggr] W_{\lambda \left(0, \, \left|F-\frac12\right|\right)}(qr) \\ 
+\sum\limits_{l=1}^{\infty}\left[C_{\lambda\left[l, \, s\left(F-\frac12\right)\right]}^{(\vee)(s)}\left(qr_0,\sqrt{q^2\sin^2\xi+m^2\cos^2\xi} \, r_0\right) W_{\lambda\left[l, \, s\left(F-\frac12\right)\right]}(qr) \right. \\ 
\left.\Biggl.-C_{\lambda\left[l, \, s\left(\frac12-F\right)\right]}^{(\wedge)(s)}\left(qr_0,\sqrt{q^2\sin^2 \xi+m^2\cos^2\xi} \, r_0\right) W_{\lambda\left[l, \, s\left(\frac12-F\right)\right]}(qr)\right]
\Biggr\},
\end{multline}
\begin{multline}\label{b12}
\left.B_{I}^{(s)}(r)\right|_{F\neq 1/2, \, \ln|\tau_s|^s=\pm\infty}=\frac{e\nu m}{(2\pi)^{2}r}\left\{\frac{1}{2\pi}\int\limits_{0}^{\infty}\frac{du}{\cosh^2(u/2)}K_1\left[2mr\cosh(u/2)\right]\Omega_{\mp}(u) \right. \\
\left. +\frac{1}{\nu}\sum\limits_{p=1}^{\left[\!\left| {\nu}/2 \right|\!\right]}K_1\left[2mr\sin(p\pi/\nu)\right]\frac{\sin[(2F-1)p\pi]}{\sin^2(p\pi/\nu)}-\frac{1}{4N}K_1(2mr)\sin[(2F-1)N\pi]\delta_{\nu,2N}\right\}\\ 
\pm\frac{e \nu r}{4 \pi^2}\int\limits_{m}^{\infty}dq q^2\left\{\frac{I_{\lambda\left[0, \, \mp\left(F-\frac12\right)\right]}(qr_0)}{K_{\lambda\left[0, \, \mp\left(F-\frac12\right)\right]}(qr_0)} W_{\lambda \left(0, \, \left|F-\frac12\right|\right)}(qr)\right.\\
\left.+\sum\limits_{l=1}^{\infty}\left[\frac{I_{\lambda \left(l, \, F-\frac12 \right)-\frac 12\mp\frac 12}(qr_0)}{K_{\lambda \left(l, \, F-\frac12 \right)-\frac 12\mp \frac 12}(qr_0)}W_{\lambda \left(l, \, F-\frac12 \right)}(qr) +\frac{I_{\lambda \left(l, \, \frac12-F \right)-\frac 12\pm\frac 12}(qr_0)}{K_{\lambda \left(l, \, \frac12-F \right)-\frac 12\pm \frac 12}(qr_0)}W_{\lambda \left(l, \, \frac12-F \right)}(qr)\right]\right\},
\end{multline}
and
\begin{multline}\label{b13}
\left.B_{I}^{(s)}(r)\right|_{F=1/2}= \frac{s e \nu m}{(2\pi)^2} \, \frac{{\rm sgn}\left(1-\tau_{s}^2\right)}{|\tau_{s}^{-1}+\tau_{s}|}\Biggl[\frac{1}{r}K_1\left(\frac{4m(r-r_0)}{|\tau_{s}^{-1}+\tau_{s}|}\right) \Biggr. \\
\left. - \int\limits_{r}^{\infty}\frac{dr'}{{r'}^2}K_1\left(\frac{4m(r'-r_0)}{|\tau_{s}^{-1}+\tau_{s}|}\right)\right] + \frac{s e \nu}{4 \pi^3}(\tau_{s}^{-1}-\tau_s)\int\limits_{m}^{\infty}dq q\Biggl\{ {\rm e}^{2qr_0} \Gamma(0,2qr)\Biggl[K\left(1-m^2/q^2\right)\Biggr.\Biggr.\\ 
\left. - \left(\frac{1+\tau_s^2}{1-\tau_s^2}\right)^2 \Pi\left(-\frac{4(1-m^2/q^2)}{(\tau_{s}^{-1}-\tau_s)^2}; 1 - m^2/q^2\right)\right]\\
\left. - 2 q r \int\limits_{0}^{\pi/2}d\xi\sum\limits_{l=1}^{\infty}C_{\nu l+\frac 12}^{(s)}\left(qr_0,\sqrt{q^2\cos^2\xi+m^2\sin^2\xi} \, r_0\right) W_{\nu l+\frac{1}{2}}(qr) \right\}.
\end{multline}
In particular,
\begin{equation}\label{b14}
\left.B_{I}^{(s)}(r)\right|_{F=1/2, \, \tau_s=1}=0
\end{equation}
and
\begin{equation}\label{b15}
\left.\lim\limits_{(-\tau_s)^s\rightarrow 1_{\mp}}B_{I}^{(s)}(r)\right|_{F=1/2}=\pm\frac{e \nu m}{(2\pi)^2}\left\{\frac{1}{r}K_1[2m(r-r_0)]-\int\limits_{r}^{\infty}\frac{dr'}{{r'}^2} K_{1}[2m(r'-r_0)]\right\}.
\end{equation}

In the case of the vanishing string tension, we obtain 
\begin{multline}\label{b16}
\left.j_{\varphi}^{(s)}(r)\right|_{\nu=1, \, F\neq 1/2, \, (\tau_s)^{-s{\rm sgn}\left(F-\frac{1}{2}\right)}\neq 0}\\ 
=\frac{m^2}{2 \pi^3}{\rm sgn}\left(F-\frac{1}{2}\right)\sin(F\pi)\int\limits_{1}^{\infty}\frac{dv}{v^2\sqrt{v^2-1}}K_2(2mrv)\cosh\left[(|2F-1|-1){\rm arccosh} v\right] \\ 
+\frac{sr}{\pi^3}\int\limits_{0}^{\pi/2}d\xi\int\limits_{m}^{\infty}dqq^2
\left\{\Biggl[\Theta\!\left(sF-\frac{s}{2}\right)
C^{(\vee)(s)}_{\frac12+s\left(F-\frac12\right)}\left(qr_0,\sqrt{q^2\sin^2\xi+m^2\cos^2\xi} \, r_0\right)\Biggr.\right. \\ 
\Biggl. - \Theta\!\left(\frac{s}{2}-sF\right) 
C^{(\wedge)(s)}_{\frac12-s\left(F-\frac12\right)}\left(qr_0,\sqrt{q^2\sin^2\xi+m^2\cos^2\xi} \, r_0\right)\Biggr] K_{F}(qr)K_{1 - F}(qr)\\
+\sum\limits_{l=1}^{\infty}\left[C_{l+s\left(F-\frac12\right)+\frac12}^{(\vee)(s)}\left(qr_0,\sqrt{q^2\sin^2\xi+m^2\cos^2\xi} \, r_0\right)K_{l+s\left(F-\frac12\right)+\frac12}(qr)K_{l+s\left(F-\frac12\right)-\frac12}(qr)\right.\\ 
\Biggl.\left.-C_{l-s\left(F-\frac12\right)+\frac12}^{(\wedge)(s)}\left(qr_0,\sqrt{q^2\sin^2\xi+m^2\cos^2\xi} \, r_0\right)K_{l-s\left(F-\frac12\right)+\frac12}(qr)K_{l-s\left(F-\frac12\right)-\frac12}(qr)\right]\Biggr\},
\end{multline}
\begin{multline}\label{b17}
\left.j_{\varphi}^{(s)}(r)\right|_{\nu=1, \, F\neq 1/2, \, \ln|\tau_s|^s=\pm\infty}\\ 
=\mp\frac{m^2}{2 \pi^3}\sin(F\pi)\int\limits_{1}^{\infty}\frac{dv}{v^2\sqrt{v^2-1}}K_2(2mrv)\cosh\left[(2F-1\pm 1){\rm arccosh} v\right]\\ 
\mp\frac{r}{2 \pi^2}\int\limits_{m}^{\infty}dq q^2
\left\{\frac{I_{\frac{1}{2}\mp\left(F-\frac{1}{2}\right)}(qr_0)}{K_{\frac{1}{2}\mp\left(F-\frac{1}{2}\right)}(qr_0)}K_{F}(qr)K_{1-F}(qr)\right.\\ 
\left.+\sum\limits_{l=1}^{\infty}\left[\frac{I_{l+F-\frac{1}{2}\mp\frac{1}{2}}(qr_0)}{K_{l+F-\frac{1}{2}\mp\frac{1}{2}}(qr_0)}K_{l+F}(qr)K_{l-1+F}(qr)+\frac{I_{l-F+\frac{1}{2}\pm\frac{1}{2}}(qr_0)}{K_{l-F+\frac{1}{2}\pm\frac{1}{2}}(qr_0)}K_{l+1-F}(qr)K_{l-F}(qr)\right]\right\},
\end{multline}
\begin{multline}\label{b18}
\left. B_{I}^{(s)}(r)\right|_{\nu=1, \, F\neq 1/2, \, (\tau_s)^{-s{\rm sgn}\left(F-\frac{1}{2}\right)}\neq 0}\\ 
=\frac{e m}{4 \pi^3 r}{\rm sgn}\left(F-\frac{1}{2}\right)\sin(F\pi)\int\limits_{1}^{\infty}\frac{dv}{v^3\sqrt{v^2-1}}K_1(2mrv)\cosh\left[(|2F-1|-1){\rm arccosh} v\right]\\ 
- \frac{s e r}{2\pi^3}\int\limits_{0}^{\pi/2}d\xi\int\limits_{m}^{\infty}dqq^2\left\{\Biggl[\Theta\!\left(sF-\frac{s}{2}\right)
C^{(\vee)(s)}_{\frac12+s\left(F-\frac12\right)}\left(qr_0,\sqrt{q^2\sin^2\xi+m^2\cos^2\xi} \, r_0\right)\Biggr.\right. \\ 
\Biggl. - \Theta\!\left(\frac{s}{2}-sF\right) 
C^{(\wedge)(s)}_{\frac12-s\left(F-\frac12\right)}\left(qr_0,\sqrt{q^2\sin^2\xi+m^2\cos^2\xi} \, r_0\right)\Biggr] W_{\frac12+|F-\frac12|}(qr)\\ 
\left.+\sum\limits_{l=1}^{\infty}\left[C_{l+s\left(F-\frac{1}{2}\right)+\frac{1}{2}}^{(\vee)(s)}\left(qr_0,\sqrt{q^2\sin^2\xi+m^2\cos^2\xi} \, r_0\right) W_{l+s\left(F-\frac{1}{2}\right)+\frac{1}{2}}(qr)\right.\right.\\ 
\Biggl.\left. - C_{l-s\left(F-\frac{1}{2}\right)+\frac{1}{2}}^{(\wedge)(s)}\left(qr_0,\sqrt{q^2\sin^2 \xi+m^2\cos^2\xi} \, r_0\right) W_{l-s\left(F-\frac{1}{2}\right)+\frac{1}{2}}(qr) \right]\Biggr\},
\end{multline}
and
\begin{multline}\label{b19}
\left.B_{I}^{(s)}(r)\right|_{\nu=1, \, F\neq 1/2, \, \ln|\tau_s|^{s}=\pm\infty}\\ 
=\mp\frac{e m}{4 \pi^{3}r}\sin(F\pi)\int\limits_{1}^{\infty}\frac{dv}{v^{3}\sqrt{v^2-1}}K_1(2mrv)\cosh[(2F-1 \pm 1){\rm arccosh} v]\\ 
\pm\frac{e r}{4 \pi^{2}}\int\limits_{m}^{\infty}dq q^2 \left\{\frac{I_{\frac{1}{2}\mp\left(F-\frac{1}{2}\right)}(qr_0)}{K_{\frac{1}{2}\mp\left(F-\frac{1}{2}\right)}(qr_0)}W_{\frac12+\left|F-\frac12\right|}(qr) \right.\\ 
\Biggl.+\sum\limits_{l=1}^{\infty}\left[\frac{I_{l+F-\frac{1}{2}\mp\frac{1}{2}}(qr_0)}{K_{l+F-\frac{1}{2}\mp\frac{1}{2}}(qr_0)} W_{l+F}(qr) + \frac{I_{l-F+\frac{1}{2}\pm\frac{1}{2}}(qr_0)}{K_{l-F+\frac{1}{2}\pm\frac{1}{2}}(qr_0)} W_{l+1-F}(qr)\right]\Biggr\};
\end{multline}
expressions at $F=1/2$ are immediately obtained by putting $\nu = 1$ in \eqref{b7} and \eqref{b13}.

\setcounter{equation}{0}
\renewcommand{\theequation}{C.\arabic{equation}}
\section*{Appendix C: Case of the infinitely thin cosmic string}

Coefficients $C^{(\wedge)(s)}_\rho$ \eqref{4.5} and $C^{(\vee)(s)}_\rho$ \eqref{4.6} obey relation
\begin{equation}\label{c1}
\lim_{r_0\rightarrow0}\,C^{(\wedge)(s)}_\rho = \lim_{r_0\rightarrow0}\,C^{(\vee)(s)}_\rho = 0, \quad \rho > 1.
\end{equation}
The nonpeculiar modes turn into regular ones at $r_0 \rightarrow 0$, as was already noted, and they are given by \eqref{a1} and \eqref{a2} with (see \eqref{a7} and \eqref{a8})
\begin{equation}\label{c2}
\lim_{r_0\rightarrow0}\,\mu^{(\wedge)(s)}_\rho = \pi/2, \quad \lim_{r_0\rightarrow0}\,\mu^{(\vee)(s)}_\rho = \pi/2.
\end{equation}
If the peculiar mode is absent, i.e. at $\nu > 1$ and $\frac{1}{2\nu} < |F-\frac12| < \frac12$, then, owing to the vanishing of $C^{(\wedge)(s)}_\rho$  and $C^{(\vee)(s)}_\rho$ in the $r_0\rightarrow0$ limit, the current and the magnetic field strength in the $r_0=0$ case are obtained by omitting sums over $l$ in  \eqref{4.12}-\eqref{4.15}.
If the peculiar mode for each value of $s$ is present, see \eqref{a5}, i.e. either at $\nu \geq 1$ and $|F-\frac12| < \frac{1}{2\nu}$, or at $\frac12 \leq \nu <1$ and $|F-\frac12| < 1 - \frac{1}{2\nu}$, then, imposing the condition of minimal irregularity on this mode in the $r_0\rightarrow0$ limit, we get (see \eqref{a9})
\begin{equation}\label{c3}
\lim_{r_0\rightarrow 0}\mu^{(c)(s)}_{\frac12-s\nu\left(F-\frac12\right)} =\left\{
\begin{array}{l}
\frac\pi2, \quad s\left(F-\frac12\right) < 0, \\
\vphantom{\int\limits_0^0}
{\rm sgn}(E)\arctan\left(\tau_s \sqrt{
\frac{1-m_3/E}{1+m_3/E}}
\right), \quad F=\frac12,\\
0,\quad s\left(F-\frac12\right) > 0.
\end{array}
\right.
\end{equation}
In the following we imply that namely \eqref{c2} and \eqref{c3} hold for the case of the infinitely thin cosmic string; then, as was already stated in the end of Section 4, we get
$$ \left.j_\varphi(r)\right|_{r_0=0} = \sum_{s=\pm 1} j_\varphi^{(a)(s)}(r) \quad {\rm and} \quad
 \left.B_I(r)\right|_{r_0=0} = \sum_{s=\pm 1} B_I^{(a)(s)}(r)
$$
with $j_\varphi^{(a)(s)}$ and $B_I^{(a)(s)}$ defined according to \eqref{4.18} and \eqref{4.19}, respectively.

In the case of $F \neq 1/2$, we use \eqref{b1} and \eqref{b11}, and note that terms in the integrals over $q$ vanish in the $r_0\rightarrow0$ limit. Summing the remaining terms over $s$, we obtain
\begin{multline}\label{c4}
\left.j_\varphi(r)\right|_{r_0=0, \, F\neq 1/2} = \frac{m^2}{\pi^2}\left\{\frac{1}{2\pi}\int\limits_0^\infty
\frac{du}{\cosh(u/2)}K_2\left[2mr \cosh(u/2)\right] \,  \Omega_{{\rm sgn}\left(F-\frac12\right)}(u)\right. \\
\left. + \frac{1}{\nu}\sum_{p=1}^{\left[\!\left| {\nu}/2 \right|\!\right]} K_2\left[2mr\sin(p\pi/\nu)\right] \,\frac{\sin[(2F-1)p\pi]}{\sin(p\pi/\nu)}
 - \frac{1}{4N} K_2\left(2mr\right) 
\sin[(2F-1)N\pi]\, \delta_{\nu, \, 2N}\right\}
\end{multline}
and
\begin{multline}\label{c5}
\left.B_I(r)\right|_{r_0=0, \, F\neq 1/2} = \frac{e \nu m}{2 \pi^2 r}\left\{ \frac{1}{2\pi}\int\limits_0^\infty
\frac{du}{\cosh^2(u/2)}K_1\left[2mr \cosh(u/2)\right] \,  \Omega_{{\rm sgn}\left(F-\frac12\right)}(u)\right. \\
\left. + \frac{1}{\nu}\sum_{p=1}^{\left[\!\left| {\nu}/2 \right|\!\right]} K_1\left[2mr\sin(p\pi/\nu)\right] \,\frac{\sin[(2F-1)p\pi]}{\sin^2(p\pi/\nu)}
 - \frac{1}{4N} K_1\left(2mr\right) 
\sin[(2F-1)N\pi]\, \delta_{\nu, \, 2N}\right\},
\end{multline}
where $\Omega_{\pm 1}(u)$ is given by \eqref{5.10}; certainly the same expressions are valid at $\nu > 1$ and 
$\frac{1}{2\nu} < |F-\frac12| < \frac12$. In the case of the vanishing string tension, the expressions are simplified:
\begin{multline}\label{c6}
\left.j_\varphi(r)\right|_{r_0=0, \, \nu=1, \, F\neq 1/2} = \frac{m^2}{\pi^3}\,{\rm sgn}\left(F-\frac12\right)\sin(F\pi)\int\limits_1^\infty \frac{dv}{v^2\sqrt{v^2-1}} \, K_2\left(2mrv\right) \\
\times\cosh\left[\left(|2F - 1| -1 \right){\rm arccosh} 
v \right] 
\end{multline}
and
\begin{multline}\label{c7}
\left.B_I(r)\right|_{r_0=0, \, \nu=1, \, F\neq 1/2} = \frac{e \nu m}{2\pi^3 r}\,{\rm sgn}\left(F-\frac12\right)\sin(F\pi)\int\limits_1^\infty \frac{dv}{v^3\sqrt{v^2-1}} \, K_1\left(2mrv\right) \\
\times\cosh\left[\left(|2F - 1| -1 \right){\rm arccosh} 
v \right]. 
\end{multline}

Turning to the case of $F = 1/2$, we foremost note that a bound state with energy $E_{BS}^{(s)} = \frac{\tau_s - \tau_s^{-1}}{\tau_s + \tau_s^{-1}} m_3$ in the gap between the continuums emerges in section $z={\rm const}$ at $\tau_s < 0$ in addition to the continuous spectrum,
\begin{equation}\label{c8}
\left( \begin{array}{c} f_{n_{\rm c}^{(s)}}^{(BS)(s)} \vphantom{\int\limits_0^0}\\
g_{n_{\rm c}^{(s)}}^{(BS)(s)}
\end{array}
\right) =  - \frac{1}{\tau_s + \tau_s^{-1}} \sqrt{\frac{2 \nu m_3}{\pi r}}  
\left( \begin{array}{c} \sqrt{-\tau_s} \\  
\sqrt{-\tau_s^{-1}} \end{array} \right)
{\rm exp}\left(\frac{2m_3 r}{\tau_s + \tau_s^{-1}}\right).
\end{equation}
We use \eqref{b7} and \eqref{b13}, note that terms in the sums over $l$ vanish in the $r_0\rightarrow0$ limit due to the vanishing of ${C}_{\nu l+\frac12}^{(s)}$ \eqref{b4}, and take this limit in the remaining terms. Summing the contribution of all modes over $s$, we obtain
\begin{multline}\label{c9}
\left.j_{\varphi}(r)\right|_{r_0 = 0, \, F=1/2}=\frac{1}{\pi^2} \sum_{s = \pm 1} s \Biggl\{{\rm sgn}\left(1-\tau_{s}^2\right)\frac{m^2}{(\tau_{s}^{-1}+\tau_{s})^2}\left[\frac{|\tau_{s}^{-1}+\tau_{s}|}{4m r}K_1\left(\frac{4m r}{|\tau_{s}^{-1}+\tau_{s}|}\right) \right.\Biggr. \\
\left. + K_0\left(\frac{4m r}{|\tau_{s}^{-1}+\tau_{s}|}\right)\right] + \frac{1}{4\pi}(\tau_{s}^{-1}-\tau_{s})\int\limits_{m}^{\infty}dq q {\rm e}^{- 2 q r}\Biggl[K\left(1-m^2/q^2\right)\Biggr.\\ 
\left.\left. - \left(\frac{1+\tau_s^2}{1-\tau_s^2}\right)^2 \Pi\left(-\frac{4(1-m^2/q^2)}{(\tau_{s}^{-1}-\tau_s)^2}; 1 - m^2/q^2\right)\right]\right\}
\end{multline}
and
\begin{multline}\label{c10}
\left.B_{I}(r)\right|_{r_0 = 0, \, F=1/2}= \frac{e \nu}{(2\pi)^2} \sum_{s = \pm 1} s \Biggl\{ {\rm sgn}\left(1-\tau_{s}^2\right)\frac{m}{|\tau_{s}^{-1}+\tau_{s}|}\left[\frac{1}{r}K_1\left(\frac{4m r}{|\tau_{s}^{-1}+\tau_{s}|}\right) \right.\Biggr. \\
\left. - \int\limits_{r}^{\infty}\frac{dr'}{{r'}^2}K_1\left(\frac{4m r'}{|\tau_{s}^{-1}+\tau_{s}|}\right)\right] + \frac{1}{\pi}(\tau_{s}^{-1}-\tau_s)\int\limits_{m}^{\infty}dq q \Gamma(0, 2qr)\Biggl[K\left(1-m^2/q^2\right)\Biggr.\\ 
\Biggl.\left. - \left(\frac{1+\tau_s^2}{1-\tau_s^2}\right)^2 \Pi\left(-\frac{4(1-m^2/q^2)}{(\tau_{s}^{-1}-\tau_s)^2}; 1 - m^2/q^2\right)\right]\Biggr\},
\end{multline}
where $K(m)$ and $
\Pi(n; m)$ are the complete elliptic integrals of the first and third kinds.
Note that \eqref{c9} and \eqref{c10}, in contrast to \eqref{c4} and \eqref{c5}, depend on $\tau_s$ and hence on $\varphi$ and $z$ in general.

Thus, it is of no surprise that the results in the case of the infinitely thin cosmic string are discontinuous at $F = 1/2$, 
\footnote{This is distinct from the case of the quantum charged scalar matter field under the Dirichlet boundary condition, when the current and the magnetic field strength that are induced in the vacuum by the infinitely thin cosmic string are continuous and vanishing at $F=1/2$, see  \cite{SiB1,SiB2,SiV9}.} see \eqref{4.20} and \eqref{4.21},
\begin{equation}\label{c11}
\left.\lim_{F\rightarrow (1/2)_{\pm}}
j_\varphi(r)\right|_{r_0=0, \, F\neq 1/2}\!=
\pm \frac{1}{2 (2\pi)^{2}r^2} \left(1+2mr\right){\rm e}^{-2mr}
\end{equation}
and
\begin{equation}\label{c12}
\left.\lim_{F\rightarrow (1/2)_{\pm}}
B_I(r)\right|_{r_0=0, \, F\neq 1/2}\!=
\pm \frac{e \nu}{(4\pi)^{2}r^2} \left[\left(1+2mr\right){\rm e}^{-2mr} - (2mr)^2 \Gamma(0,2mr)\right];
\end{equation}
the discontinuity is independent of $\nu$ for 
$\left.j_\varphi(r)\right|_{r_0=0, \, F\neq 1/2}$ and is linear in $\nu$ for $\left.B_I(r)\right|_{r_0=0, \, F\neq 1/2}$. 
Note also that $\left.j^{(s)}_\varphi(r)\right|_{r_0=0, \, F = 1/2}$ and $\left.B^{(s)}_I(r)\right|_{r_0=0, \, F = 1/2}$, vanishing at $\tau_s=1$, increase monotonically in absolute value as $\tau_s$ departs from this value. Their maximal absolute values for $\tau_s > 0$ are reached at $\tau_s^{-1} = 0$ or $\tau_s = 0$, being equal to
\begin{equation}\label{c13}
\lim_{\ln |\tau_s |^s \rightarrow \mp \infty}\left.j^{(s)}_\varphi(r)\right|_{r_0=0, \, F = 1/2} \!=\left.\lim_{F\rightarrow (1/2)_{\pm}}
j^{(s)}_\varphi(r)\right|_{r_0=0, F\neq 1/2}\!=
\pm \frac{1}{(4\pi)^{2}r^2} \left(1+2mr\right){\rm e}^{-2mr}
\end{equation}
and
\begin{multline}\label{c14}
\lim_{\ln |\tau_s |^s \rightarrow \mp \infty}\left.B^{(s)}_I(r)\right|_{r_0=0, \, F = 1/2}\!=
\left.\lim_{F\rightarrow (1/2)_{\pm}}
B^{(s)}_I(r)\right|_{r_0=0, F\neq 1/2} \\ 
= \pm \frac{e \nu}{2 (4\pi)^{2}r^2}  \left[\left(1+2mr\right){\rm e}^{-2mr} - (2mr)^2 \Gamma(0,2mr)\right].
\end{multline}
The increase in absolute value persists further for 
$\tau_s < 0$, reaching the maximum at $\tau_s=-1$,
\begin{equation}\label{c15}
\left.\lim_{(-\tau_s)^s \rightarrow 1_{\mp}}
j^{(s)}_\varphi(r)\right|_{r_0=0, \, F = 1/2}\!=
\pm \frac{m}{(2\pi)^{2}r} \left[K_1\left(2mr\right) + 2mr K_0\left(2mr\right)\right]
\end{equation}
and 
\begin{equation}\label{c16}
\left.\lim_{(-\tau_s)^s \rightarrow 1_{\mp}}
B^{(s)}_I(r)\right|_{r_0=0, \, F = 1/2}\!=
\pm \frac{e \nu m}{2 (2\pi)^{2}r} \left[K_1\left(2mr\right) + 2mr \int\limits_r^\infty \frac{dr'}{r'}  K_0\left(2mr'\right)\right].     
\end{equation}

Turning now to the temporal component of the induced vacuum current in the $r_0 = 0$ case, we note that, as a consequence of \eqref{c1}, it vanishes at $F \neq 1/2$. Since quantity
$$
\lim_{r_0\rightarrow 0}\left({\tilde C}^{(\vee)}_{1/2} - {\tilde C}^{(\wedge)}_{1/2}\right) = \lim_{r_0\rightarrow 0}\left({C}^{(\vee)}_{1/2} - {C}^{(\wedge)}_{1/2}\right) 
$$
is nonvanishing, we obtain
\begin{multline}\label{c17}
\left.j^{0}(r)\right|_{r_0 = 0, \, F=1/2} = 
\frac{\nu}{(2\pi)^2 r} \sum_{s = \pm 1} 
\Biggl\{\frac{\tau_{s}^{-1}-\tau_{s}}{8 r^2}
\left(1 +2 m r\right) \, {\rm e}^{-2m r} \Biggr. \\
 - \frac{m}{2 r}{\rm sgn}\left(\tau_{s}^{-1}-\tau_{s}\right)\left[K_1\left(\frac{4m r}{|\tau_{s}^{-1}+\tau_{s}|}\right) + \frac{4m r}{|\tau_{s}^{-1}+\tau_{s}|}K_0\left(\frac{4m r}{|\tau_{s}^{-1}+\tau_{s}|}\right)\right] \\
 - (\tau_{s}^{-2}-\tau_{s}^2)\int\limits_{m}^{\infty}dq q {\rm e}^{-2q r}\Biggl[K\left(1-m^2/q^2\right)\Biggr.\\ 
\Biggl.\left. - \left(\frac{1+\tau_s^2}{1-\tau_s^2}\right)^2 \Pi\left(-\frac{4(1-m^2/q^2)}{(\tau_{s}^{-1}-\tau_s)^2}; 1 - m^2/q^2\right)\right]\Biggr\}.
\end{multline}
Note also relations [cf. \eqref{b9}, \eqref{c13}, and \eqref{c15}]
\begin{equation}\label{c18}
\left.j^{0 (s)}(r)\right|_{F = 1/2, \, \tau_s =1}\!= 0,
\end{equation}
\begin{equation}\label{c19}
\left.\lim_{\ln |\tau_s |^s \rightarrow \mp \infty}
j^{0 (s)}(r)\right|_{r_0=0, \, F = 1/2}\!=
\pm \frac{s \nu}{2 \pi^{3}r} \int\limits_{m}^{\infty}dq q {\rm e}^{- 2 q r} E\left(1-m^2/q^2\right),
\end{equation}
where
$$
E(m)=\int\limits_{0}^{1}dt \sqrt{\frac{1-m t^2}{1-t^2}}
$$
is the complete elliptic integral of the second kind, see \cite{Abra}, and
\begin{equation}\label{c20}
\left.\lim_{(-\tau_s)^s \rightarrow 1_{\mp}}
j^{0 (s)}(r)\right|_{r_0=0, \, F = 1/2}\!=
\pm \frac{s \nu m}{(2\pi)^{2}r^2} \left[K_1\left(2mr\right) + 2mr K_0\left(2mr\right)\right].
\end{equation}


\begin{thebibliography}{0}    

\bibitem{Ki1} T.W.B. Kibble,   J. Phys. A: Math. Gen. \textbf{9}, 1387 (1976).

\bibitem{Ki2}
T.W.B. Kibble,  Phys. Rep. \textbf{67}, 183 (1980).

\bibitem{Zel}%
Ya.B. Zeldovich,  Mon. Not. Roy. Astron. Soc. \textbf{192}, 663
(1980).

\bibitem{Vil1}%
A. Vilenkin, Phys. Rev. D \textbf{23}, 852 (1981).

\bibitem{Vil2}%
A. Vilenkin, Phys. Rev. D \textbf{24}, 2082 (1981).

\bibitem{Vil3}%
A. Vilenkin and E.P.S. Shellard, {\it Cosmic Strings and Other Topological Defects} (Cambridge Univ. Press, Cambridge UK, 1994)

\bibitem{Ki3}%
M.B. Hindmarsh and T.W.B. Kibble, Rep. Progr. Phys.
\textbf{58}, 477 (1995).

\bibitem{Dam}%
T. Damour and A. Vilenkin, Phys. Rev. D \textbf{71}, 063510 (2005).

\bibitem{Ber}%
V. Berezinsky, B. Hnatyk, and A. Vilenkin, Phys. Rev. D
\textbf{64}, 043004 (2001).

\bibitem{Bhat}%
P. Bhattaharjee and G. Sigl. Phys. Rep. \textbf{327}, 109 (2000).

\bibitem{Sar}%
S. Sarangi and S.H.H. Tye, Phys. Lett. B \textbf{536}, 185 (2002).

\bibitem{Jean}%
R. Jeannerot, J. Rocher, and M. Sakellariadou, Phys. Rev. D \textbf{68}, 103514 (2003).

\bibitem{Dva}%
G. Dvali and A. Vilenkin, J. Cosmol. Astropart. Phys. JCAP \textbf{0403}, 010 (2004).

\bibitem{Pol}%
J. Polchinski, Int. J. Mod. Phys. A \textbf{20}, 3413 (2005).

\bibitem{Sak}%
M. Sakellariadou, Nucl. Phys. Proc. Suppl. \textbf{192-193}, 68 (2009).

\bibitem{Cop}%
E.J. Copeland and T.W.B. Kibble, Proc. Roy. Soc. London A \textbf{466}, 623 (2010).

\bibitem{Bat}%
R.A. Battye, B. Garbrecht, A. Moss, and H. Stoica, J. Cosmol.
Astropart. Phys. JCAP \textbf{0801}, 020 (2008).

\bibitem{Vol}%
G.E. Volovik, JETP Lett. {\bf 67}, 698 (1998) [Pisma Zh. Eksp. Teor. Fiz. {\bf 67}, 666 (1998)].

\bibitem{Ge}
A.K. Geim and K.S. Novoselov, Nature Mater. {\bf 6}, 183 (2007).

\bibitem{Cah}
 S. Cahangirov, M. Topsakal, E. Akturk,  H. Sahin, and S. Ciraci, Phys. Rev. Lett. {\bf 102}, 236804 (2009).

\bibitem{Liu}
H. Liu, A.T. Neal, Z. Zhu, Z. Luo, X. Xu, D. Tomanek, and P.D. Ye, ACS Nano {\bf 8}, 4033 (2014).

\bibitem{Tsu}
C.C. Tsuei and J.R. Kirtley, Rev. Mod. Phys.  {\bf 72}, 969 (2000).

\bibitem{Qi}
X.L. Qi and S.C. Zhang, Rev. Mod. Phys. {\bf 83}, 1057 (2011).

\bibitem{Ehre}%
W. Ehrenberg and R.E. Siday, Proc. Phys. Soc. London B \textbf{62}, 8 (1949).

\bibitem{Aha}%
Y. Aharonov and D. Bohm, Phys. Rev. {\bf 115}, 485 (1959).

\bibitem{Gor}%
P. Gornicki, Ann. Phys. (N.Y.) {\bf 202}, 271 (1990).

\bibitem{Si0}%
Yu.A. Sitenko, Nucl. Phys. B {\bf 342}, 655 (1990).

\bibitem{Flek}%
E.G. Flekkoy and J.M. Leinaas, Intern. J. Mod. Phys. A {\bf 06}, 5327 (1991).

\bibitem{Par}%
R.R. Parwani and A.S. Goldhaber, Nucl. Phys. B {\bf 359}, 483 (1991).

\bibitem{Sit6}%
Yu.A. Sitenko, Phys. Lett. B \textbf{387},  334 (1996).

\bibitem{Sit9}%
Yu.A. Sitenko, Phys. Atom. Nucl. \textbf{62},  1056 (1999).

\bibitem{Bez1}
E.R. Bezerra de Mello, V. Bezerra, A.A. Saharian, and V.M. Bardeghyan, Phys. Rev. D {\bf 82}, 085033 (2010).

\bibitem{Bel1}
S. Bellucci, E.R. Bezerra de Mello, and A.A. Saharian, Phys. Rev. D {\bf 83}, 085017 (2011).

\bibitem{Bez2}
E.R. Bezerra de Mello, F. Moraes, and A.A. Saharian, Phys. Rev. D {\bf 85}, 045016 (2012).

\bibitem{Bel2}
S. Bellucci, E.R. Bezerra de Mello, A. de Padua, and A.A. Saharian, Eur. Phys. J. C {\bf 74}, 2688 (2014).

\bibitem{Bez4}
M.M. de Sousa, R. Ribeiro, and E.R. Bezerra de Mello, Phys. Rev. D {\bf 93}, 043545 (2016).

\bibitem{Bez5}
M.M. de Sousa, R. Ribeiro, and E.R. Bezerra de Mello, Phys. Rev. D {\bf 95}, 045005 (2017).

\bibitem{SiG}%
Yu.A. Sitenko and V.M. Gorkavenko, Phys. Rev. D \textbf{100}, 085011 (2019).

\bibitem{Bee}
A.R. Akhmerov and C.W.J. Beenakker, Phys. Rev. B {\bf 77}, 085423 (2008).

\bibitem{Wie}
M.H. Al-Hashimi and U.-J. Wiese, Ann. Phys. (Amsterdam) {\bf 327}, 1 (2012).

\bibitem{Si1}
Yu.A. Sitenko, Phys. Rev. D {\bf 91}, 085012 (2015).

\bibitem{SiY}
Yu.A. Sitenko and S.A. Yushchenko, Intern. J. Mod. Phys. A \textbf{30}, 1550184 (2015).

\bibitem{Nero}
 A. Neronov and I. Vovk, Science {\bf 328}, 73 (2010).
 
\bibitem{Subr}
K. Subramanian, Rep. Progr. Phys. {\bf 79}, 076901 (2016).

\bibitem{Bog}%
P.N. Bogolioubov, Ann. Inst. Henri Poincare A \textbf{8}, 163 (1968).

\bibitem{Cho1}
A. Chodos, R.L. Jaffe, K. Johnson, C.B. Thorn, and V. Weisskopf, Phys. Rev. D {\bf 9}, 3471 (1974).

\bibitem{Joh}
K. Johnson, Acta Phys. Pol. B {\bf 6}, 865 (1975).

\bibitem{Ree}%
M. Reed and B. Simon, {\it Methods of Modern Mathematical Physics II.
Fourier Analysis, Self-Adjointness} (Academic Press, New York, 1975).

\bibitem{Alb}%
S. Albeverio, F. Gesztezy, R.
Hoegh-Krohn, and H. Holden, {\it Solvable Models in Quantum Mechanics},
(Springer-Verlag, Berlin, 1988).

\bibitem{Si3}%
Yu.A. Sitenko, J. Phys.: Conf. Series \textbf{670}, 012048 (2016).

\bibitem{Gor2}
V.M. Gorkavenko, Yu.A. Sitenko, and O.B. Stepanov, J. Phys. A: Math. Theor. {\bf 43}, 175401 (2010).

\bibitem{Gor3}
V.M. Gorkavenko, Yu.A. Sitenko, and O.B. Stepanov, Intern. J. Mod. Phys. A {\bf 26}, 3889 (2011).

\bibitem{Gor4}
V.M. Gorkavenko, Yu.A. Sitenko, and O.B. Stepanov, Intern. J. Mod. Phys. A {\bf 28}, 1350161 (2013).

\bibitem{Gor1}
V.M. Gorkavenko, I.V. Ivanchenko, and Yu.A. Sitenko, Intern. J. Mod. Phys. A {\bf 31}, 1650017 (2016).

\bibitem{SiR} Yu.A. Sitenko and D.G. Rakityansky, Phys. Atom. Nucl. \textbf{60},  1497 (1997).

\bibitem{Si7} Yu.A. Sitenko, Phys. Atom. Nucl. \textbf{60},  2102 (1997);               (E) \textbf{62}, 1084 (1999) .

\bibitem{Si9}
Yu.A. Sitenko, Phys. Atom. Nucl. \textbf{62},  1767 (1999).

\bibitem{SiV7}
Yu.A. Sitenko and N.D. Vlasii, Nucl. Phys. B {\bf 787}, 241 (2007).

\bibitem{SiV1}%
Yu.A. Sitenko and N.D. Vlasii, J. Phys.: Conf. Series \textbf{129},
012008 (2008).

\bibitem{SiV2}%
Yu.A. Sitenko and N.D. Vlasii, Low Temp. Phys. \textbf{34}, 826 (2008) [Fiz. Nizk. Temp. \textbf{34}, 1049 (2008)].

\bibitem{Si18}%
Yu.A. Sitenko and V.M. Gorkavenko, Low Temp. Phys. \textbf{44}, 1261 (2018) [Fiz. Nizk. Temp. \textbf{44}, 1618 (2018)].

\bibitem{Abra}
{\it Handbook of Mathematical
Functions}, edited by M. Abramowitz and I.A. Stegun (Dover, New York, 1972).

\bibitem{SiB1}%
Yu.A. Sitenko and A.Yu. Babansky, Mod. Phys. Lett. A \textbf{13}, 379 (1998).

\bibitem{SiB2}%
Yu.A. Sitenko and A.Yu. Babansky, Phys. At. Nucl. \textbf{61}, 1594 (1998).

\bibitem{SiV9} Yu.A. Sitenko and N.D. Vlasii, Class. Quantum Grav. \textbf{26}, 195009 (2009).


\end{thebibliography}
\end{document}